\newcounter{myctr}

\documentclass[12pt]{article}
\usepackage[a4paper,top=3cm, bottom=3cm, left=3cm, right=3cm]{geometry}

\usepackage[small]{caption}

\usepackage{url}
\usepackage{graphicx}
\usepackage{rotating}

\begin{document}

\makeatletter
\def\@biblabel#1{[#1]}
\makeatother

%
%

%

\renewcommand{\topfraction}{0.85}
\renewcommand{\textfraction}{0.1}


\newcommand{\half}{\frac{1}{2}}
\newcommand{\bea}{\begin{eqnarray}}
\newcommand{\eea}{\end{eqnarray}}
\newcommand{\beq}{\begin{equation}}
\newcommand{\eeq}{\end{equation}}
\newcommand{\nnel}{\nonumber \\ {}}

\newcommand{\tseref}[1]{(\ref{#1})}
\newcommand{\tref}[1]{(\ref{#1})}


\renewcommand{\thefootnote}{\arabic{footnote}}
\newcommand{\tnote}[1]{} 
\newcommand{\tsepreprint}[1]{#1}
\newcommand{\tseurl}[2]{\texttt{#1}}
\newcommand{\tsetitle}[1]{\emph{#1}}

\newcommand{\tsevec}[1]{\mathbf{#1}}
\newcommand{\tsemat}[1]{{\mathbf{\textsf{#1}}}}
\newcommand{\tsematg}[1]{{\boldsymbol{#1}}}

\newcommand{\Amat}{\tsemat{A}}
\newcommand{\Bmat}{\tsemat{B}}
\newcommand{\Btildemat}{\tilde{\Bmat}}
\newcommand{\Btildeinmat}{\tilde{\Bmat}^\mathrm{in}}
\newcommand{\Btildeoutmat}{\tilde{\Bmat}^\mathrm{out}}
\newcommand{\Cmat}{\tsemat{C}}
\newcommand{\Ctildemat}{\tilde{\tsemat{C}}}
\newcommand{\Dmat}{\tsemat{D}}
\newcommand{\Dtildemat}{\tilde{\tsemat{D}}}
\newcommand{\Emat}{\tsemat{E}}
\newcommand{\Etildemat}{\tilde{\tsemat{E}}}
\newcommand{\Fmat}{\tsemat{F}}

%
%

\markboth{Evans, Rivers and Knappett}{Interactions in space for archaeological models}

%
%

\renewcommand{\thefootnote}{\fnsymbol{footnote}}
\begin{center}
{\Large Interactions In Space For Archaeological Models\footnote{Contribution to special issue of Advances in Complex Systems from the conference `Cultural Evolution in Spatially Structured Populations', UCL September 2010. \texttt{Imperial/TP/11/TSE/1}}}\\[0.5cm]
T.S.\ Evans, R.J.\ Rivers \\[6pt]
Theoretical Physics, Imperial College London, \\
London, SW7 2AZ, U.K. \\[6pt]
C.\ Knappett \\[6pt]
Department of Art
Sidney Smith Hall, Room 6036
100 St George Street,\\
University of Toronto
Toronto
Ontario
M5S 3G3
\end{center}



\begin{abstract}
In this article we examine a variety of quantitative models for describing archaeological networks, with particular emphasis on the maritime networks of the Aegean Middle Bronze Age.
In particular, we discriminate between those gravitational networks that are most likely (maximum entropy) and most efficient (best cost/benefit outcomes).
\end{abstract}

\renewcommand{\thefootnote}{\arabic{footnote}}
\setcounter{footnote}{0}

\section{Introduction}

In archaeology, the primary sources of information are usually finds from specific sites.  In order to understand the social, cultural and political context of these finds it is essential that we understand how sites relate.    However, it often requires so much effort to obtain the physical information from a single site that there is a danger that we become `site-centric'.
Despite our best attempts at deducing relationships from the artefacts found at them, there is often little direct information about how sites interact. Quantitative modelling provides one response to this challenge.
Good quantitative modelling can be insensitive to poor data, make assumptions and biases clear and debatable, and can allow us to provide possible answers to questions that could not be asked any other way.  In the best case, such answers can be checked later from the archaeological record.

One of the major problems that we attempt to address is to determine how the relationships between (archaeological) sites are conditioned by geographical space; to what extent does the exchange between them transcend their geographical constraints?
As the basis for our modelling we have found the language of complex networks particularly useful as a network is \emph{both} a set of vertices ---  the sites --- \emph{and} a set of edges, or links --- the relationships between them. Socio-spatial networks do not arise and develop arbitrarily. Their maintenance incurs explicit and implicit `costs', both in sustaining sites and in supporting links \cite{B10}. The common agency in the networks that we shall discuss below is the need to regulate or distribute these costs. We now encounter a conundrum. Despite these general features, the modelling of social networks requires very specific tailoring for individual case studies. This is contingent on the level of social organisation, the nature of the exchange, relevant distance scales, travel technology, etc. of the society being studied. Thus, for exemplary purposes, we have found it useful to work with one particular data set, namely the Middle Bronze Age (MBA) to Late Bronze Age I (LB I) S.Aegean. Nonetheless, we shall try to be as general as possible within this framework.

\section{Networks}

As we have said, networks primarily consist of vertices (or nodes), representing agents or populations or resources, and edges (or links), which represent the exchange between them, from the physical trade of goods to the transmission of ideas or culture. The MBA Aegean has several distinct features that lends itself to a network analysis.  There is a relatively clear temporal delineation both at the beginning of the MBA and at the end  of LB I.  From the start of the MBA (c.2000 BC), society on Crete, often dubbed `Minoan', sees  a series of innovations, such as monumental architecture at palatial centres, new craft technologies
 and more sustained long-distance exchange. This last trait is most likely enabled by the innovation of the sail that seems to have occurred around this time, supplementing what was previously a paddle-based maritime technology. The end of our period of study, LB I, c.1450 BC, sees the destruction of the Minoan palaces and a swing in power and influence to the Greek mainland; this marks a useful delineation for analytical purposes. The physical boundaries of our focus for study are also relatively well defined, though long range trade beyond the Aegean was essential e.g. in providing tin, and questions regarding the relationship to Egypt remain important.  Nevertheless, as a first approximation, limiting our analysis to the Aegean region provides a good example of where modelling may be of particular use.  There is a large amount of material from this era, with Akrotiri and Minoan palaces such as Knossos being rich sources.
In this paper we will describe our network model, \texttt{ariadne}, and show how it is related to other approaches to archaeological modelling, using these MBA maritime exchange networks as a testing ground.
In so doing we will go beyond the qualitative descriptions given in our previous work \cite{EKR07,KER08}.

\subsection{Defining Our Vertices}\label{svertices}

One of the main ways in which the nature of society impacts on the construction of networks is in determining  what constitutes vertices which, in principle, could vary in scale from individuals/households to communities to islands.   Ironically this also reflects the fact that most network analysis is also site-centric (although graph theory does provide some tools to get around this \cite{EL09,E10}).
In general, island archipelagos are ideal for networks as the geography provides a natural definition for our vertices. Our working hypothesis is that the  detailed behaviour of individuals can be subsumed in the behaviour of the 'island' community. In this regard, we have taken each vertex to correspond to an island, or a large part of an island or, in the case of N. Crete and other coastal areas, to an isolated centre of population. The latter sites effectively behave as islands because of the difficulty of land travel.
This hypothesis is contingent on a maritime technology permitting long-distance travel that enables the island (or `island-like' coastal site) to be thought of as the basic unit for `trade' and has been discussed in detail elsewhere \cite{EKR07,KER08}. However, once  we move away from individuals to communities, it is apparent that each vertex has two very different attributes; its resource base and the exploitation of these resources. The former,  the carrying capacity of the vertex, is part of our {\it input}.  The latter is represented by the `size'  of the community that inhabits it.  That is part of our {\it output}.

In many models there is no discrimination between carrying capacity and population (i.e. one is taken proportional to the other) and, often, the carrying capacities of the sites are set equal.  This may well be appropriate for some cases such as pre-urban civilisations where land was occupied in small villages of roughly equal size.  The work of Broodbank \cite{B00} on the Early Bronze Age in the Cyclades is a good example of this approach.  There the potential cultivable land of each island was assessed and the density of sites on habitable land was chosen to be uniform.
However we are interested in an era when significant differences emerge between the size of different sites.  These differences reflect more than just local differences in resources and we are looking to understand how interactions with other sites may have supported very large sites.  Thus it is important that site sizes are variable in our model, both in terms of the fixed inputs given and in terms of the outputs, since we expect a complex interdependence of interactions and site sizes.
We have chosen a set of major known islands or coastal sites as representing the most important locations for vertices. See Figure 1, with details in Table \ref{t39sites}. The choice is made on the basis of the archaeological record. It is sufficient for our purposes in Table 1 to classify the carrying capacities of the sites as `small' (S), `medium' (M) or  `large' (L). What this means quantitatively will be explained later.  In so doing we are assuming that other locations are peripheral to the dynamics of the whole system or at least their effect is well represented by the dominant site in their region.

\begin{figure}[htb]
\centering
\includegraphics[width=0.9\textwidth]{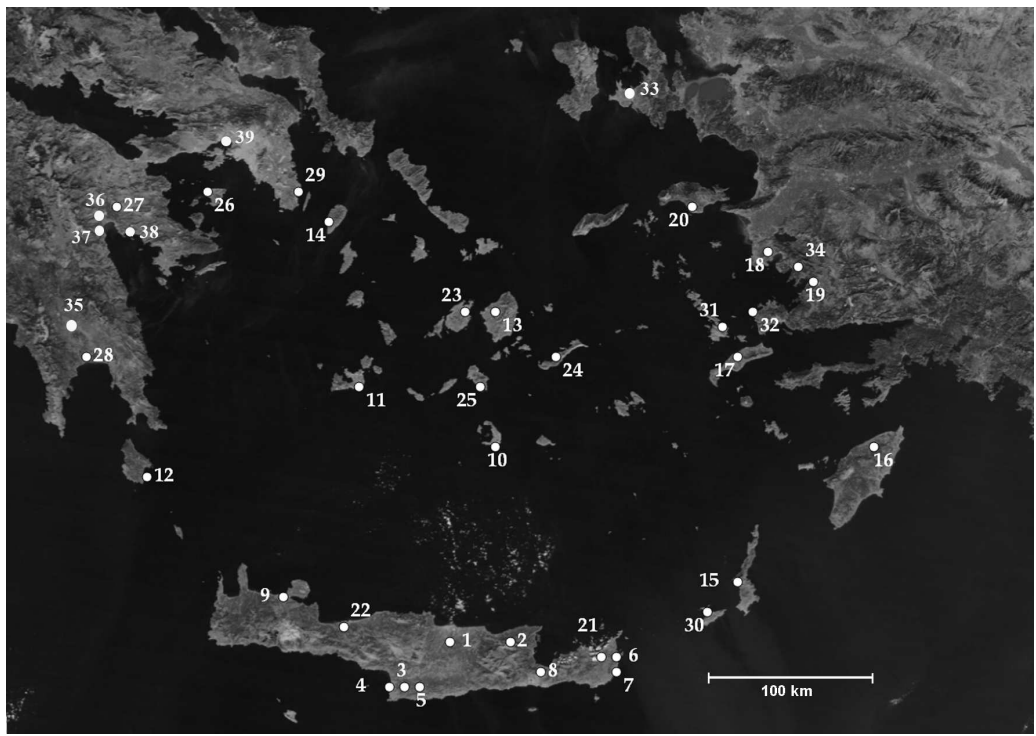}
\caption{Thirty nine important Aegean sites of the MBA (Middle Bronze Age).}\label{f39sites}
\end{figure}

\begin{table}[htbp]
\centering
\begin{tabular}{|r@{ }l@{ }l|r@{ }l@{ }l|r@{ }l@{ }l|}
\hline
1. & Knossos  & (L) & 14. &      Kea  & (M) & 27. &      Mycenae  & (L) \\ \hline
2. & Malia  & (L) & 15. &      Karpathos  & (S) & 28. &      Ayios Stephanos  & (L) \\ \hline
3. & Phaistos  & (L) & 16. &      Rhodes  & (L) & 29. &      Lavrion  & (M) \\ \hline
4. & Kommos  & (M) & 17. &      Kos  & (M) & 30. &      Kasos  & (S) \\ \hline
5. & Ayia Triadha  & (L) & 18. &      Miletus  & (L) & 31. &      Kalymnos  & (S) \\ \hline
6. & Palaikastro  & (L) & 19. &      Iasos  & (M) & 32. &      Myndus  & (M) \\ \hline
7. & Zakros  & (M) & 20. &      Samos  & (M) & 33. &      Cesme  & (M) \\ \hline
8. & Gournia  & (L) & 21. &      Petras  & (L) & 34. &      Akbuk  & (M) \\ \hline
9. & Chania  & (L) & 22. &      Rethymnon  & (L) & 35. &      Menelaion  & (S) \\ \hline
10. & Thera  & (M) & 23. &      Paroikia  & (M) & 36. &      Argos  & (M) \\ \hline
11. & Phylakopi  & (M) & 24. &      Amorgos  & (S) & 37. &      Lerna  & (M) \\ \hline
12. & Kastri  & (M) & 25. &      Ios  & (S) & 38. &      Asine  & (S) \\ \hline
13. & Naxos  & (L) & 26. &      Aegina  & (M) & 39. &      Eleusis  & (M) \\ \hline
\end{tabular}
\caption{Thirty-nine sites of the MBA Aegean. Note Knossos as site 1 and Akrotiri, subsequently destroyed in the eruption of Thera, as site 10.  The S,M,L indicates a rough assignment of size --- small, medium or large.}
\label{t39sites}
\end{table}

As for notation,  we denote vertices using lowercase mid Latin indices $i,j, ...$. The fixed carrying capacities of sites will be denoted $S_i$ (habitable land available). In some models the total population of a site may differ from this as we will indicate.

\subsection{Defining our Edges}

The edges represent the interactions and can be seen within the framework of different types of space e.g. artefact space \cite{Sind07,S07b,T10}.
 However we take the view that archaeology is rooted in geographical space and it is an unavoidable constraint on interactions.

This means that one of the most important inputs to our model will be a table of distances between our sites.  These may be simple Euclidean (as-the-crow-flies) distances.  A more sophisticated approach will use estimates of typical journey times. This will be sensitive to the technology available and could require information on currents, typical winds, slopes, and even security.  For instance in our examples we have analysed the layout of islands in the Aegean by hand to take promontories into account, but not the effects of winds and currents.  Likewise for land travel we picked routes by hand that reflected current three-dimensional geography, arguing that the accuracy of a full least cost path analysis
 is unnecessary given the level of approximation.  Finally in the examples shown here, we choose to penalise land travel over sea travel by a `friction' coefficient of 3.0. Results are largely insensitive to reasonable values of the coefficient (larger than unity), primarily affecting the width of Crete. We will not look at this issue in any depth here, rather we will take it that these effective distances between sites are given.  The issue we address in detail is how to model the actual interactions, given a framework of sites and their separations.

For notation, if the attribute  of an edge from $i$ to $j$ is $A_{ij}$ then the matrix ${\mathbf{\textsf{A}}}$ of these elements is an adjacency matrix for the network. When $A_{ij}$ defines the flow from site $i$ to site $j$ it will be denoted by $F_{ij}$. Note that unless otherwise stated, models exclude self-loops excluded, that is $F_{ii}=0$.
The effective distance from site $i$ to site $j$ will be $d_{ij}$.  For all examples here, this is symmetric $d_{ij}=d_{ji}$, although as discussed above this is not always appropriate.

\section{Geographical Models}

Our emphasis, and that of the other main models that we shall discuss, is based on network analysis, as we have already indicated.
 However, before going into details we stress that there are alternative approaches that work more with `zones of influence' than exchange directly. Although vertices are preserved, networks themselves are abandoned, and we begin by discussing such models briefly.

\subsection{Geographical Models without Networks}

There are several examples of models for archaeology, based on significant sites such as we have defined, yet without any explicit edges defined and so no network.  If we treat all sites as equal we can construct partitions (or partial partitions) of geographical space into something like `zones of influence'.  This is a very  old idea, reincarnated in its simplest form as simple Voronoi, or Thiessen, polygons, each of which contain all the points in space for which the single site at the centre of that polygon is the closest site.  They can be used to indicate zones of control of each site, e.g.\ for Etruscan Cities using Euclidean distances \cite{R75}.  The XTent model of Renfrew and Level \cite{RL79} is a generalisation for cases with different site sizes. Here site $i$ of size $S_i$ is deemed to `dominate' site $j$ of size $S_j$ if $\tan(\theta) > d_{ij}/(S_i-S_j)$ where $d_{ij}$ is the distance between the two sites and $\tan(\theta)$ is a parameter of the model.  In the XTent model one can work with just the sites (rather than all points in space needed for Voronoi diagrams) and we can represent the resulting hierarchy of sites
as a directed network. An example of how this may be used with modern GIS techniques to gain a good estimate of actual walking time, rather than using simple Euclidean distance, has been given for Neopalatial Crete by Bevan \cite{B10c}.   However, such a network representation adds little and the model is not usually visualised in this way, e.g. see \cite{RL79,B10c}.

\subsection{Simple Geographical Networks}

There are two simple ways to construct a network which captures non-trivial information about the global interactions between sites, both based on a thresholding of the distance matrix.

\begin{figure}[htb]
\centering
\includegraphics[width=0.49\textwidth]{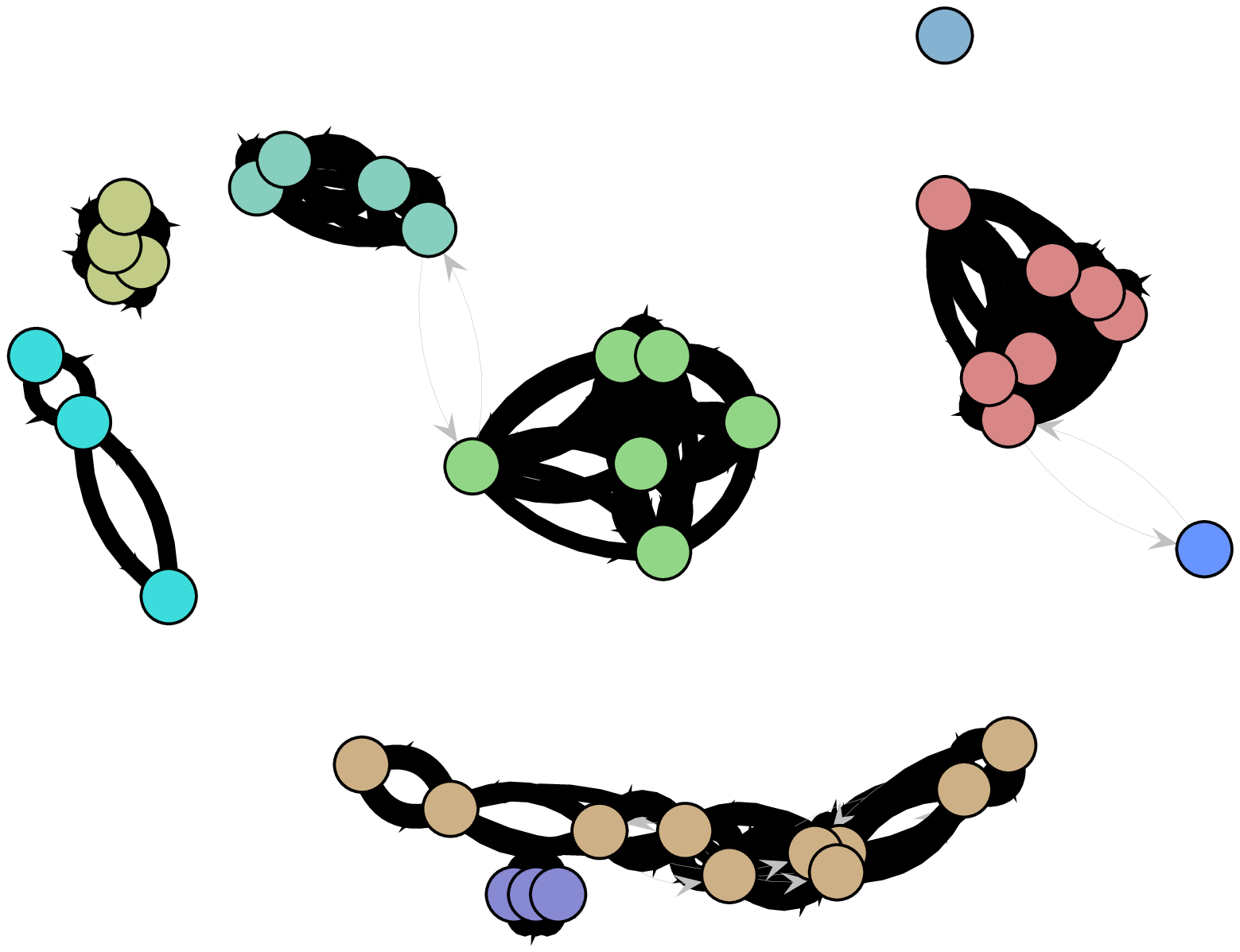}
\includegraphics[width=0.49\textwidth]{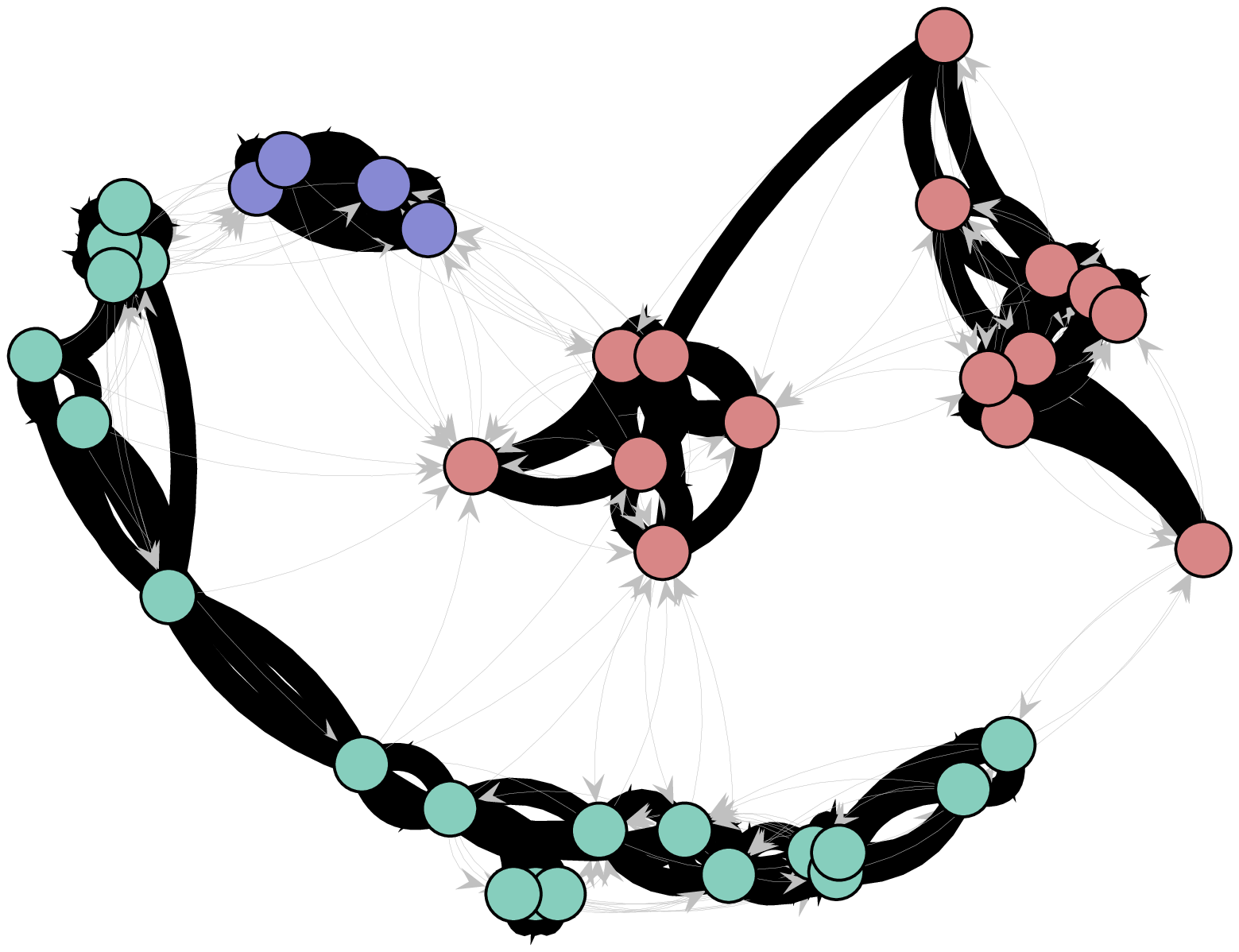}
\caption{On the left an example of an MDN (Maximum distance network) while the right hand network is a k-nearest-neighbour graph or PPA (Proximal Point Analysis) network.  These are for the 39 MBA Aegean sites of Fig.\ \ref{f39sites}, using estimated travel times where land travel has a friction coefficient of 3 compared to sea travel.  In the MDN network, thick edges in black are present if the separation is less than $D=100$km (thin grey lines indicate edges between 100 and 110km).  In the PPA network thick black lines indicate the case where each vertex is connected to its three nearest neighbours $k=3$ (thin grey lines indicate connections to fourth and fifth nearest neighbours).  In both cases the colour of vertices indicates the connectivity of the network defined by the thick black edges only.}\label{fMDN}\label{fPPA}
\end{figure}

The first is a `maximum distance network' (MDN) in which an edge from site $i$ to site $j$ is connected if the distance is less than some model parameter $D$. In the language of links, we might construe this  as imposing a link likelihood $A_{ij}= \Theta(D-d_{ij})$ for the journey, where $\Theta(x)$ is the step function, taking value $1$ for $d_{ij} <D$ and zero otherwise.   MDN is used in many fields to establish distance scales, e.g.\ as a model of ad-hoc wireless networks formed between mobile devices \cite{SH10}.  A great deal of work exists on these types of models (particularly `random geometric graphs' \cite{P03,SH10}), but they appear not to have been widely used for models in archaeology. See Fig.\ \ref{fMDN} for the Aegean when $D = 100km$ and $110km$. [The reader should impose the pattern of sites in Fig.\ \ref{fMDN} on the geographic template of Fig.\ \ref{f39sites}, to identify the sites.]  At this value the S. Aegean splits into four identifiable zones; Crete, the Cyclades, the Dodecanese and the Peloponnese. If we increase the distance to $D=130km$, these regions  connect [not shown]. These are relevant values for the MBA, for which the dominant marine technology is sail, with vessels routinely capable of travelling large distances in single journeys. As a marker we note that the distance from Knossos to Akrotiri is in excess of $100km$, which we believe was achievable in a single journey. Typically we take $D$ to be $100km$ or somewhat greater.

The second thresholding method is more common in archaeology where it is known as PPA (Proximal Point Analysis) \cite{T77,I83,HH91,B00,C07,T10}
though it has received less attention in other fields, e.g.\ as `k-nearest-neighbour graphs' \cite{PS04,BBSW05}. Here each site is connected to its $k$ nearest neighbours ($k$ is a parameter of the model) to give a directed network, see Fig.\ \ref{fPPA}, though the directions are usually ignored in the literature. An example is given in Fig.\ \ref{fPPA} for $k=3$ nearest neighbours. It differs strongly from its MDN counterpart in that, by definition, sites on the extrema of the map that are separated from their neighbours will connect nonetheless, despite the distances, because the assumption is that they will connect somehow, independent of how easy this may be.  As a result, there is a tendency for sites to be connected in `strings'.

The MDN can be thought of as imposing prohibitive costs on implementing long single journeys and PPA as imposing prohibitive costs on sustaining more than a few social interactions.
The last simple model worth mentioning is the elementary `Gravity model' \cite{ES90,OW94}.  For this the flow $F_{ij}$ from site $i$ to site $j$ is assumed to take the form $F_{ij}= S_iS_jf(d_{ij}/D)$ (with $F_{ii}=0$) where the distance `potential' $f(x)$ is a monotonic decreasing function, reflecting the ease of travel from $i$ to $j$ or, as we shall see, the effective cost of travel.  In the context of the MBA Aegean we want short trips by sea to be relatively easy, or of low cost, while we want a strong cutoff or penalty for single trips of distance $D$ or more. We stress that, unlike for the MDN, $D$ is no longer an absolute cut-off, but, with $f(1) = 0.5$, sets the scale above which single journeys become increasingly difficult. $D$ therefore takes somewhat smaller values than we would read off from the MDN.

We have chosen a generic form for $f(x)$ that is a smoothed out version of the MDN , namely
\begin{equation}
 f(x) = [1+x^{\beta_1}]^{-\beta_2} \, . \label{fdef}
\end{equation}
In our work we have used only
$\beta_1=4.0$ and $\beta_2=1.0$. As long as there is a plateau for $x < 1$ followed by rapid falloff for $x > 1$ the outputs are similar.
As we see in Fig.3, it produces a dense network, although many of the long distance links will be weak.  In particular either of the previous two thresholding techniques could be used to create a network from the flows $F_{ij}$.  In this sense this Gravity model can be seen as a generalisation of the previous two models to sites of different sizes.

\begin{figure}
\centering
\includegraphics[width=0.49\textwidth]{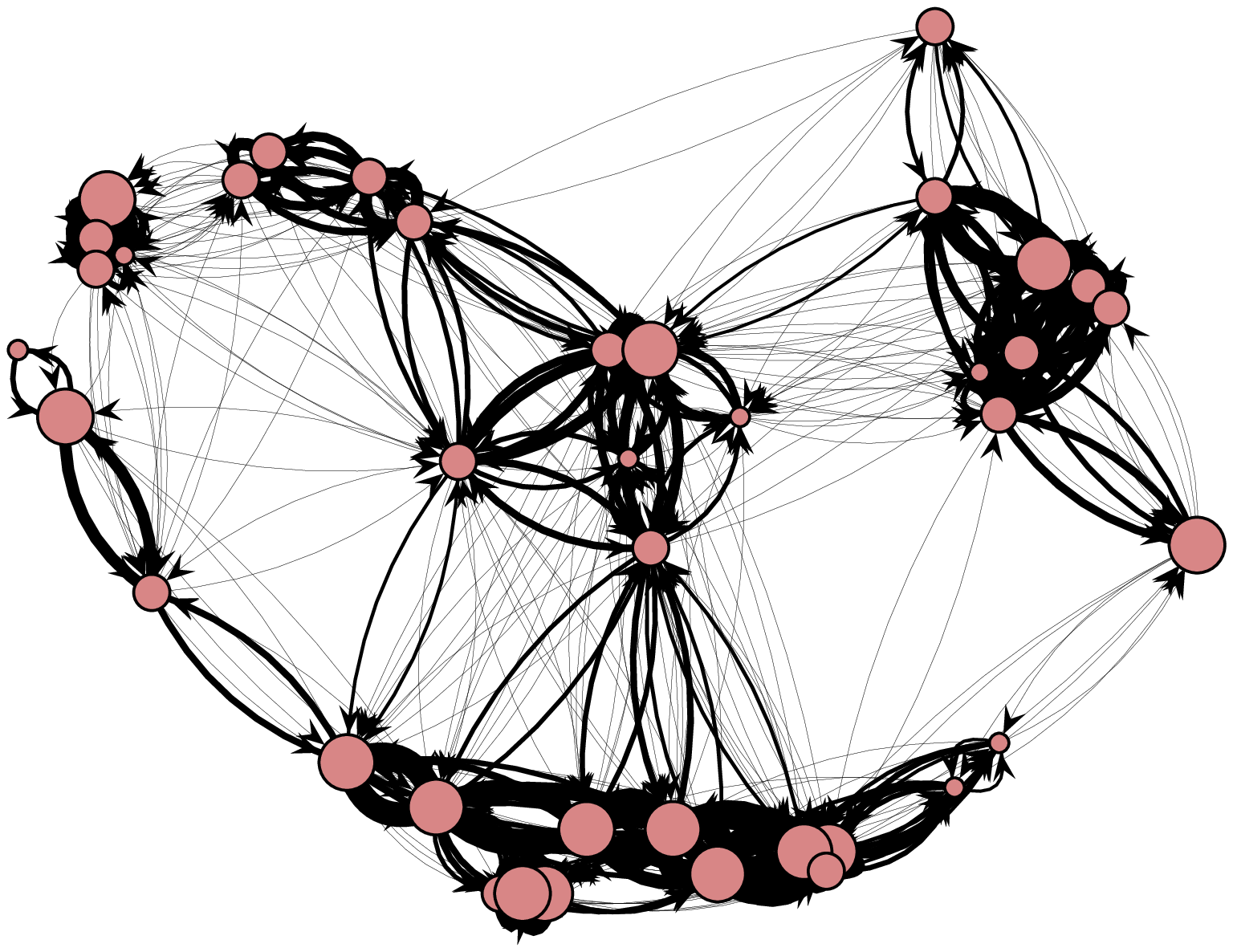}
\includegraphics[width=0.49\textwidth]{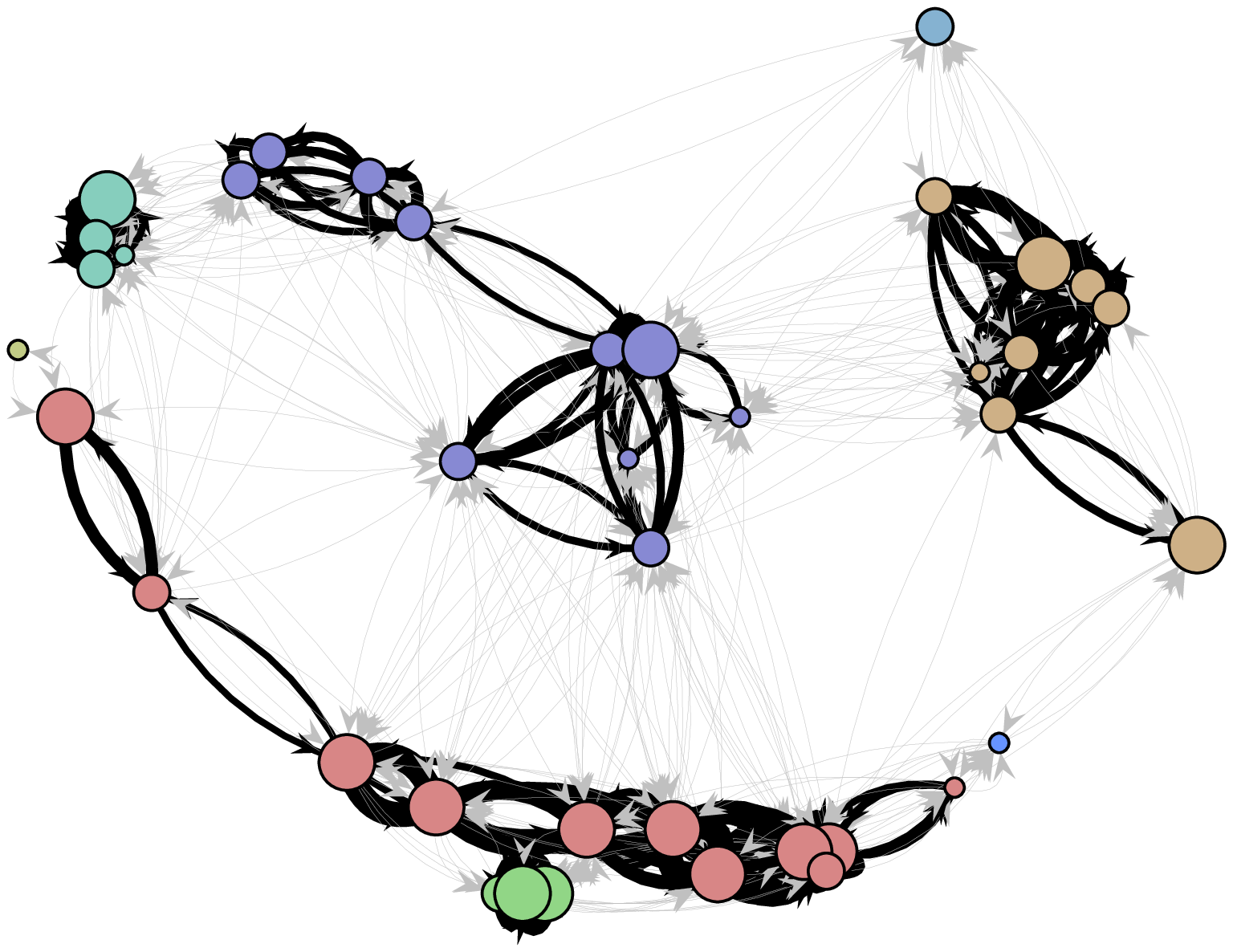}
\caption{A simple Gravity Model with edge weights $F_{ij}= S_iS_jf(d_{ij}/D)$ for the 39 MBA Aegean sites of Fig.\ \ref{f39sites}, using estimated travel times where land travel has a friction coefficient of 3 compared to sea travel. The function $f(x)$ used is given in (\ref{fdef}) with a distance scale of $D=100$km.  Vertices are proportional to the site size (0.33, 0.67 or 1.00 for small, medium and large sites).  On the right all edges are shown but with their thickness proportional to the flow $F_{ij}$. The same network is shown on the right but now all links with $F_{ij}<0.23$ are shown only as thin grey lines.  In addition the colour of vertices indicates the connectivity of the network of strong ($F_{ij}>0.23$) links.}\label{fsGM}
\end{figure}

In the first two models sites are of fixed and equal size.   This, and subsequent, Gravity Models avoid this limitation. In particular, by aggregating island communities into an island `centre of population' and a similar conflation of inter-island exchange, they assume that, at the island level, the whole is the sum of the parts. By this coarse-graining, they minimise our need to have detailed local knowledge, in that sense making the best of poor data. However, the elementary model above fails to capture the rich variety of possible interactions --- there is no feedback between interactions and site size.  In arriving at the flow from $i$ to $j$, we could change the location of all other sites and it would make no difference.

\section{Optimal Geographical Networks}

The models which are most fit for our purpose are ones in which the strength of the interactions reflect both the local geographical topology between two sites but also the wider regional structure in which these sites reside. In different ways the two model types that have been developed most fully are optimal models, either looking for the `most likely' networks, all other things being equal, or the most `efficient' networks.

The simplest example of the former is that of the {\it doubly constrained} Gravity Model (DCGM).  Here the flow takes the same form as before but with coefficients determined self-consistently:-
\begin{eqnarray}
 F_{ij} &=& A_i O_i B_j I_j f(\frac{d_{ij}}{D}) \, , \;
 \frac{1}{A_i} = \sum_k B_k I_k f(\frac{d_{ij}}{D}) \, , \;
 \frac{1}{B_j} = \sum_k A_k O_k f(\frac{d_{ij}}{D}) \, .
\label{flowDCGM}
\end{eqnarray}
Although not immediately transparent, solutions of this form are in fact turning points with respect to the $N(N-1)$ variables $F_{ij}$ (as $F_{ii}=0$) of the Hamiltonian
\begin{eqnarray}
 H &=&
   \sum_{(i,j)} F_{ij} \big(\ln(F_{ij})  -1  \big)
  -\sum_i \alpha_i  \Big[O_i-\sum_j F_{ij} \Big]
  -\sum_j \alpha_j^\prime  \Big[I_j-\sum_i F_{ij} \Big]
 \nonumber \\
 && \hspace*{6cm} -\beta \Big[C-\sum_{ij}(F_{ij}c_{ij})\Big] \, .
 \label{HdefDCGM}
\end{eqnarray}

The first term in (\ref{HdefDCGM}) is the negative of the entropy of a  network of flows $F_{ij}$.
Maximising this entropy term produces the most likely distribution of the total exchange, $\sum_{i,j}F_{ij}$, amongst all the possible edges, in the absence of any further knowledge as to how exchange occurs.  As all edges are treated equally by this entropy term, it is the remaining terms which produce a non-trivial solution.

The $N$ parameters $\{\alpha_i\}$ are Lagrange multipliers which enforce constraints on the total outflow, $O_i= \sum_j F_{ij}$, from each site $i$.  The $\{\alpha_j^\prime\}$ do the same job for the total inflow of site $j$, $I_j= \sum_i F_{ij}$. Here $O_i$ and $I_j$ are input parameters of the model which we have taken proportional to the carrying capacity/population.  The simplicity of the model allows for an algebraic solution  for these Lagrange multipliers which allows us to replace the $\{\alpha, \alpha_i^\prime\}$ with the normalisations $\{A_i\}$ and $\{B_j\}$ in (\ref{flowDCGM}).  Enforcing these constraints ensures the flow along the edge from $i$ to $j$ depends directly on the flow from $i$ to all other sites and upon the flow into $j$ from all other sites. The largest contributions will be from sites near to $i$ and/or to $j$ so the interactions do depend on the whole region.  Changing the location of other sites in the region will alter the flow from $i$ to $j$, the type of behaviour we are seeking.

The last Lagrange multiplier, $\beta$, is associated with the total `cost' $C$ for the pattern of flows.  Here the cost per unit of flow is given by $c_{ij}$ for a link from $i$ to $j$. The total cost $C$ could reflect many factors rather than monetary and so it is rarely known in a physical example.  So rather than specify the total cost as a parameter of the model, it is normal to keep $\beta$ as a parameter.  Larger (smaller) $\beta$ means higher (lower) costs.
In practice $\beta$ dependence becomes absorbed into one or more parameters into the travel, or distance, `potential' $f(x)$ of (\ref{fdef}) on making the identification $f(d_{ij}/D)= \exp (- \beta c_{ij})$.
Thus larger $D$ means cheaper costs and corresponds to larger $\beta$.

\begin{figure}[htb]
\centering
\includegraphics[width=0.49\textwidth]{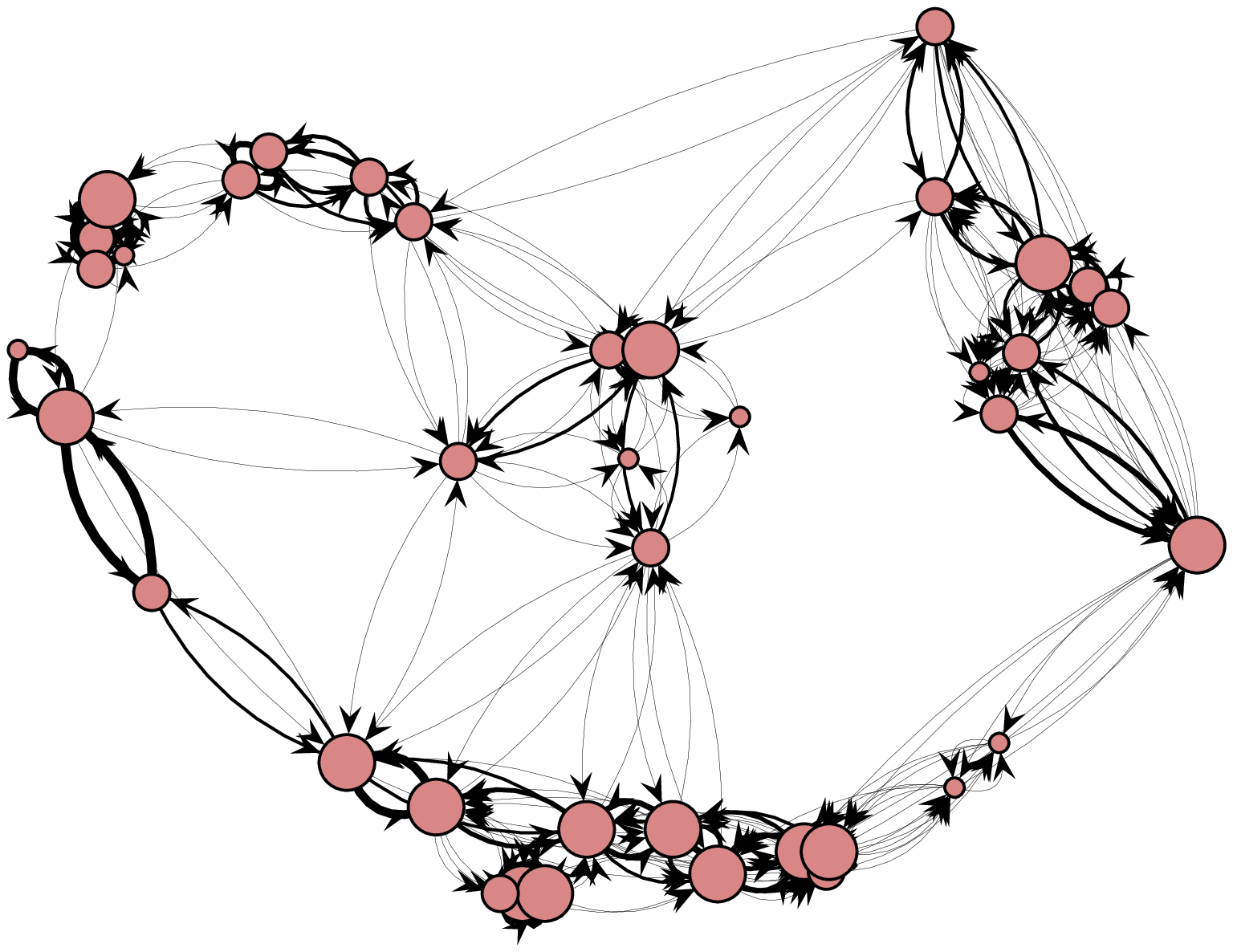}
\includegraphics[width=0.49\textwidth]{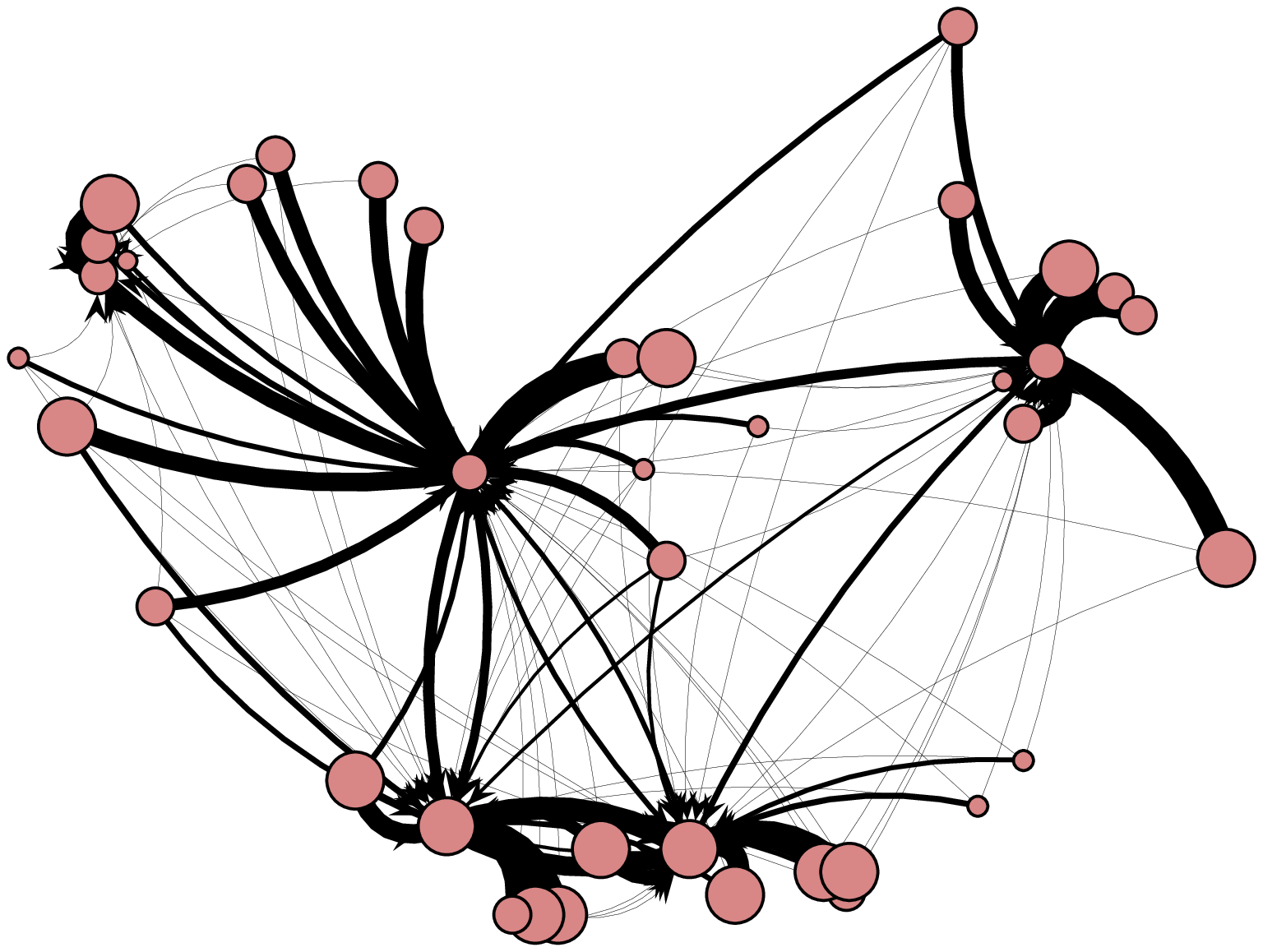}
\caption{Two different optimal Gravity models for the 39 MBA Aegean sites of Fig.\ \ref{f39sites}, using estimated travel times where land travel has a friction coefficient of 3 compared to sea travel.In both cases the distance scale is $D=100$km.  Vertices are proportional to the site size (0.33, 0.67 or 1.00 for small, medium and large sites).  On the left solutions for the doubly constrained Gravity model (DCGM) are shown. On the right is the Rihll and Wilson Gravity model (RWGM) (\ref{RWiter}) with $\gamma=1.3$. } \label{fDCGM}\label{fRWGM}
\end{figure}

An exemplary network of the DCGM is given in Fig.\ \ref{fDCGM}.
The drawback of this model is that we have to fix both input and output site sizes and these do not respond to the pattern of interactions which emerges. For example, in our MBA Aegean context,  Knossos can never alter the volume of economic, cultural or social exchange from the values we gave it at the start. Further, because inflows and outflows are fixed, we get a repeat of the pattern in the PPA that remote sites are still strongly connected, even if distances are large.

Some of these issues are addressed in the work of Rihll and Wilson \cite{RW87,RW91} who, instead, use a model originally devised to study the emergence of dominant retail centres \cite{CW85,W08,WD10,DW09,DW10}, which we now summarise.
The basic Rihll and Wilson Gravity model (RWGM) gives the flow $F_{ij}$ from site $i$ to site $j$
as
\begin{equation}
F_{ij}=A_i O_i I_j^\gamma f(d_{ij}/D) \, , \qquad  A_i^{-1} = \sum_k I_k^\gamma f(d_{ik}/D) \, .
\label{FRW}
\end{equation}
 As before, $O_i$ is the total outflow of site $i$, $O_i= \sum_j F_{ij}$ and $I_j$ is the total inflow of site $j$, $I_j= \sum_i F_{ij}$. The difference from the DCGM is that, while outflows $O_i$ are still input parameters of the theory (taken to be proportional to site carrying capacity), the inflows $I_j$ are now outputs determined by the model. These are used by Rihll and Wilson to assign an importance to a site\footnote{In a second self-consistent version, Rihll and Wilson set $O_i=I_i$. The model parameters now include an initial value for $I_i$ for each site.  We will not consider this variant further.}. Solutions to this nonlinear equation are in fact turning points of the Hamiltonian
\begin{eqnarray}
 H &=& \sum_{i,j} F_{ij} \big(\ln(F_{ij})  -1  \big) -\sum_i \alpha_i  \Big[O_i-\sum_j F_{ij} \Big] -\beta \Big[C-\sum_{i,j}(F_{ij}c_{ij})\Big]
 \nonumber \\
 && \hspace*{5cm}- \gamma \Big[X-\sum_{i,j}F_{ij} \big(\ln(\sum_k F_{kj} ) -1\big) \Big],
 \label{HdefRW}
\end{eqnarray}
again understood as maximising entropy.
As before, we have encoded $\beta$ and the cost function $c_{ij}$ in the distance potential $f$ of (\ref{fdef}). Likewise we retain the $\gamma$ Lagrange multiplier in the solution (\ref{FRW}) rather than specify the unknown $X$ in the Hamiltonian (\ref{HdefRW}).
An important solution is where the flow into one or more sites is zero so that $I_j=F_{ij}=0$ for a finite number of sites $j$, but for all values of index $i$.

In principle there are many ways to find these solutions.
Rihll and Wilson generate a sequence of values $\{I_j(0), I_j(1), \ldots ,I_j(t) , \ldots \}$ where
\begin{equation}
I_j (t+1)=\sum_i A_i(t) O_i (I_j (t))^\gamma f(d_{ij}/D ) \, , \qquad  A_i^{-1} (t) = \sum_k I_k^\gamma(t) f(d_{ik}/D)
\label{RWiter}
\end{equation}
Provided the parameters are chosen suitably the limiting value of this sequence, $\lim_{t \rightarrow \infty} I_j(t)$, generates flows which optimise $H$ of (\ref{HdefRW}).
Rihll and Wilson use an ``egalitarian hypothesis'' and set the fixed outputs and the initial inputs all equal $O_i =I_j(t=0)=1$. This is deemed equivalent to the assumption that ``all sites were approximately equal in size and importance at the beginning of the period under consideration'', since there is often little information available to make any other hypothesis.

In principle, $t$ could just represent some computational time and have no physical significance. However it is sometimes interpreted as physical time \cite{W08}  in which case this represents a whole further set of assumptions.  In particular one can view the finite difference equation as the Euler method for solving the differential equation
\begin{equation}
\frac{dI_j}{dt}= \epsilon \left(\sum_i A_i O_i (I_j (t))^\gamma e^{(-\beta c_{ij} )} -KI_j \right)
\end{equation}
where $K$ and $\epsilon$ are new constants.

An example of the RWGM is given in Fig.\ \ref{fRWGM} for $\gamma > 1$, which generates hubs. Taking values $\gamma <1$ recreates networks closer in structure to those of the DCGM (not shown).
As with the DCGM, the enforced outflows ensure that however remote a site may be, it will always be connected.  As a result, although we have not been able to show this in the limited Figures in the text, these models give networks whose important links are largely insensitive to our choice of $D$, within reason (e.g. varying $D$ from $80km$ to $130km$).

Whatever choice we make, one possibility is to exclude such peripheral sites, as we have for our chosen MBA sites and for the Iron Age Greek mainland sites chosen by Rihll and Wilson.

\section{\texttt{ariadne}}

We conclude with a discussion of the model, named \texttt{ariadne}, that we have developed over a period of time for examining MBA maritime networks in the S. Aegean \cite{EKR07,KER08}.
We wanted site sizes to be informed, but not absolutely constrained, by the local resources, with physical geography placing similar limitations on interactions.  Rather than look for the most likely networks, as we have been doing with gravity models, we are looking for the most efficient networks from the viewpoint of the costs and benefits that the network demands and provides.

In our model the input parameters are the fixed local site resources, $S_i$, and the site locations plus four independent parameters characterising the component parts of the `Hamiltonian' or `social potential' $H$, understood as representing the costs {\it minus} the benefits in sustaining the network. As outputs the model produces the flows $F_{ij}$ between different sites $i$ and $j$ but now also produces self-loop terms $F_{ii} \geq 0$ to represent parts of the population not involved in interactions.  The values are found by minimising the following Hamiltonian\footnote{In previous papers \cite{EKR07,KER08,RKE11} we worked in terms of fractional values for total site size $v_i$ and for flows $e_{ij}$ with $W_i = S_i v_i$ and $F_{ij} = S_iv_ie_{ij}$. The Hamiltonian is then
$H = - \kappa \sum_i 4 S_iv_i (1-v_i) - \lambda \sum_{i,j} (S_iv_i) e_{ij} f(d_{ij}/D) (S_jv_j) + J\sum_i S_iv_i  + \mu \sum_{i,j} S_iv_ie_{ij}$.}
$H$ (i.e. maximising benefits with respect to costs):-
  \begin{eqnarray}
H  &=&
 - \kappa \sum_i 4 W_i (1-W_i/S_i)
 - \lambda \sum_{i,j} F_{ij} f(d_{ij}/D) W_j
 + J \sum_i W_i
 + \mu \sum_{i,j} F_{ij}.
 \label{Hariadne}
\end{eqnarray}
where $W_i = \sum_j F_{ij}$ represents a variable total population at site $i$. Note that in this model $W_i$ is not the same as the total output $O_i = \sum_{j | j\neq i} F_{ij} = W_i-F_{ii}$ if some of the population at site $i$ stays at home, $F_{ii}>0$.

The first term of $H$ is proportional to $\kappa$ is a measure of the benefit of local productivity.  It is minimised by setting $W_i=S_i/2$, only over-exploitation $W_i>S_i$ produces positive energy (negative benefits).  This term is similar in form to $W_i [\ln(W_i)-1]$ which would be present if we were maximising the entropy associated with the arrangement of the total `population', $\sum_j W_j$, amongst the sites.  This is in contrast to the optimal Gravity models where it is the entropy associated with the distribution of people amongst the \emph{edges} which is maximised.

The second term, proportional to $\lambda$, provides benefits of exchange, as tempered by geography through the function $f$ of (\ref{fdef}). Links over distances much longer than the parameter $D$ produce relatively little benefit to reflect the low probability of such direct links being maintained.  The benefit of a link is also deemed to be in proportion to the target site's total size $W_j$.  As the source site's size is related to the flow from $i$ to $j$, $F_{ij}$, this term is reminiscent of the product of source and target sites sizes that characterise gravity models as in (\ref{flowDCGM}).  However our product is in the Hamiltonian, not in the solution for the flows. Our nonlinear dependence on the site sizes in this term is a key distinction between our models and the optimised gravity models presented earlier.  Thus, an increase in interactions from $i$ to $j$ will produce a positive feedback and increasing site $j$'s size will produce even greater benefits, encouraging in turn more growth in the flow between the two sites, possibly (but not necessarily) requiring an increase in the size of site $i$. The first term proportional to $\kappa$ prevents this process from running away but it does allow the volume of interactions to grow if the dynamics favours it\footnote{While one could imagine other dependencies on the target site size, say $(W_j)^\gamma$ with some new model parameter $\gamma$, we have not found such an additional parameter necessary.}.

The last two terms, proportional to $J$ and $\mu$ are the costs in sustaining the population and the total flow.  The $J$ term is equivalent to setting the total output rather than individual site outputs in Gravity models, so that $J=\sum_i \alpha_i$ in the optimal Gravity models of (\ref{HdefDCGM}) and (\ref{HdefRW}).  The term proportional to $\mu$ is different only if $F_{ii}\neq 0$ for some sites $i$, emphasising that the full potential for interactions need not be met in our model --- some `people' can stay at home and need not `trade' if there is insufficient benefit.  This is another improvement over gravity models as in \texttt{ariadne} a remote site $i$ separated by several $D$ from the nearest neighbour will probably not interact, $F_{ij}=0$ if $i \neq j$, and will probably maintain a population of about $S_i/2$ based on the benefits of local resources encoded through the $\kappa$ term.

We also impose some further constraints on our variables.  Clearly we demand that $W_i\geq 0$ and $F_{ij}\geq0$.  More importantly we also impose a short range cutoff such that, for sites separated by less than a certain minimum distance, $d_{ij}>d_\mathrm{min}$, we set $F_{ij}=0$.  Technically, if we impose this cutoff then we can split any site into two pieces separated by less than $d_\mathrm{min}$, and these two small sites behave like a single unified site.  In this way we have an explicit scale for our coarse graining.  We suggest that this scale be set by the distance of travel in a day by land, say $10$km.  In our case it only effects three sites on the Southern coasts of Crete.

We note that the scale of the Hamiltonian is irrelevant so we usually choose $\kappa=1.0$ leaving us with the three Hamiltonian parameters $\lambda,J,\mu$ (the last two may be of any sign) plus the distance scales $D$ and $d_\mathrm{min}$.  The final parameters are the fixed site resources $S_i$ along with their separations $d_{ij}$. However, unlike for the previous models, the networks are now more sensitive to the values of $D$ chosen. For $D$ noticeably less than $100km$ the networks becomes fragmented, whereas for $D$ more than $130km$, say, the networks have a tendency to become very dense. This is a direct reflection of the typical distance scale $\approx 130km$ (not quite identified with $D$, see earlier) for which the sites form a connected whole. We can invert the problem to say that the existence of a strong maritime network requires a marine technology whose single journeys match the distances required to produce a network. As we noted earlier, this is plausible, given the $100km$ plus of the distance from Knossos to Akrotiri.

\begin{figure}[htb]
\centering
\includegraphics[width=0.49\textwidth]{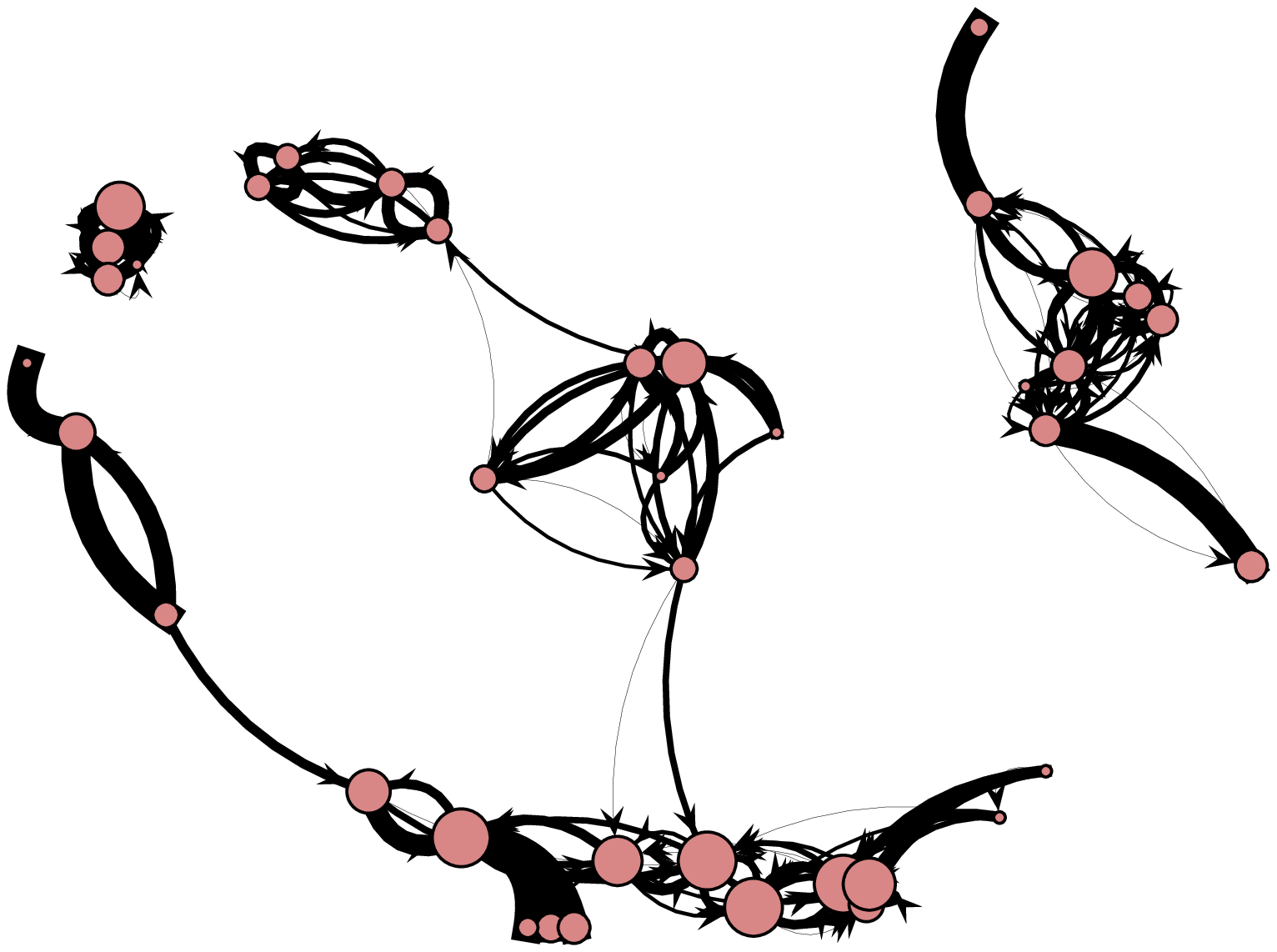}
\includegraphics[width=0.49\textwidth]{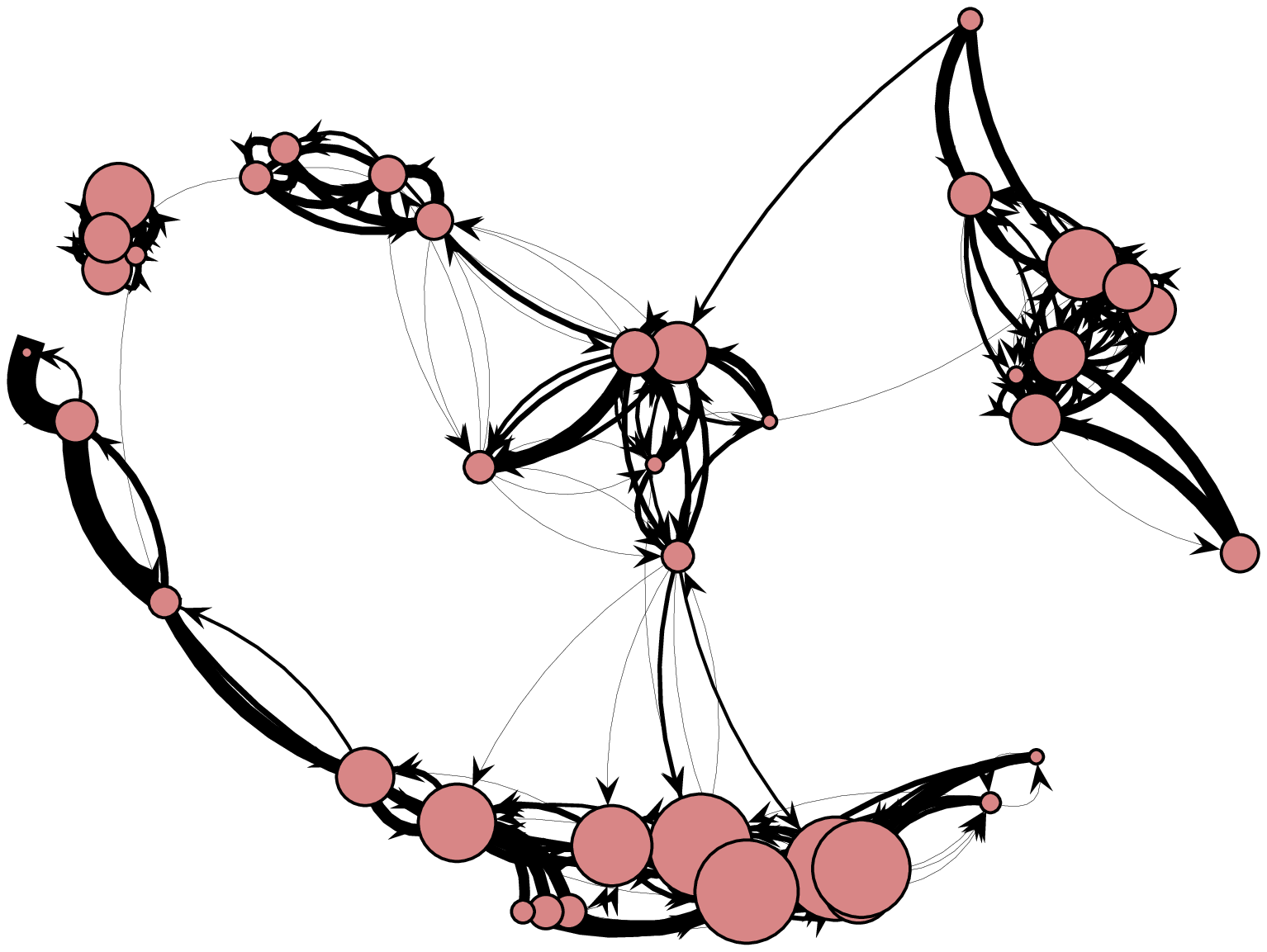}
\\
\includegraphics[width=0.49\textwidth]{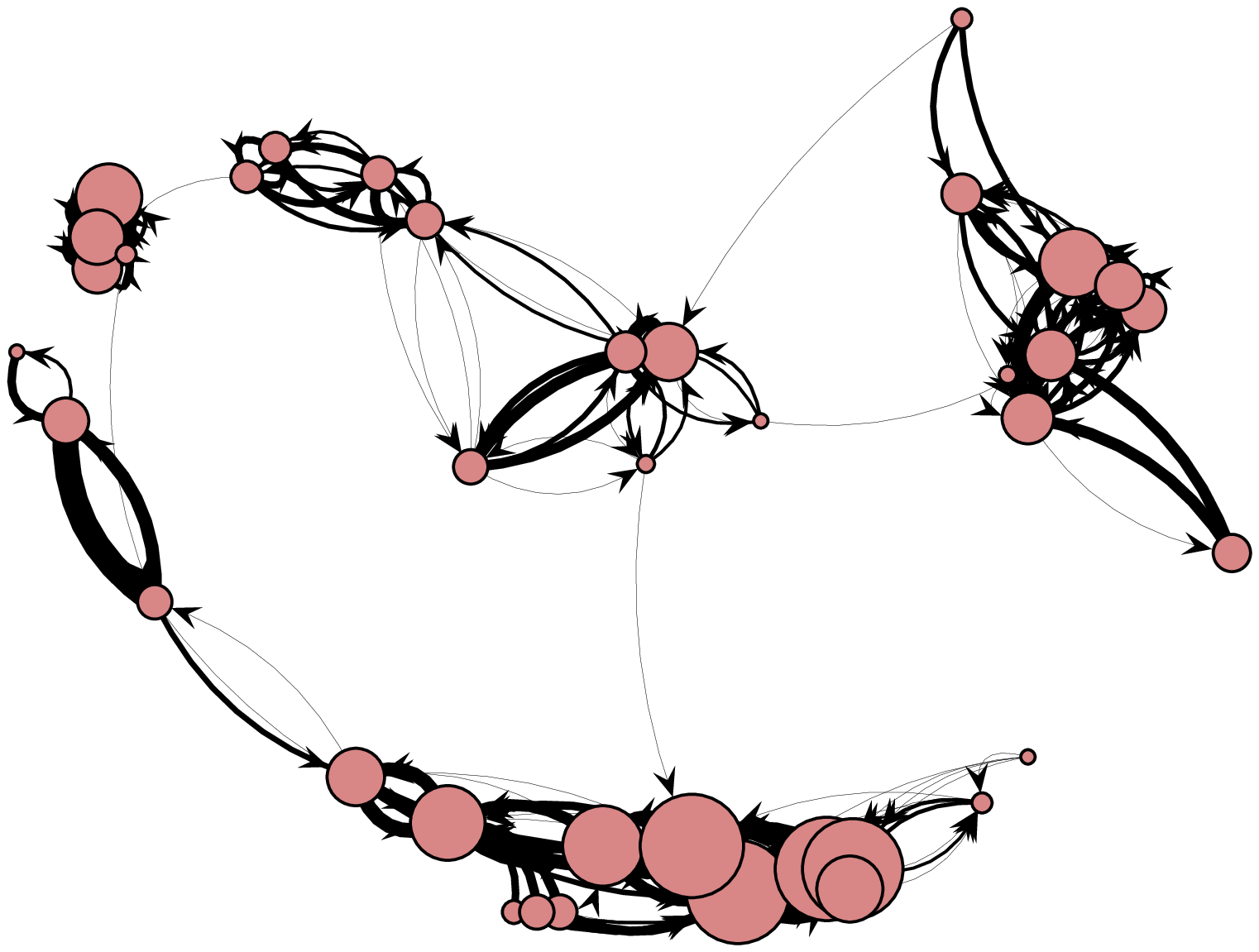}
\includegraphics[width=0.49\textwidth]{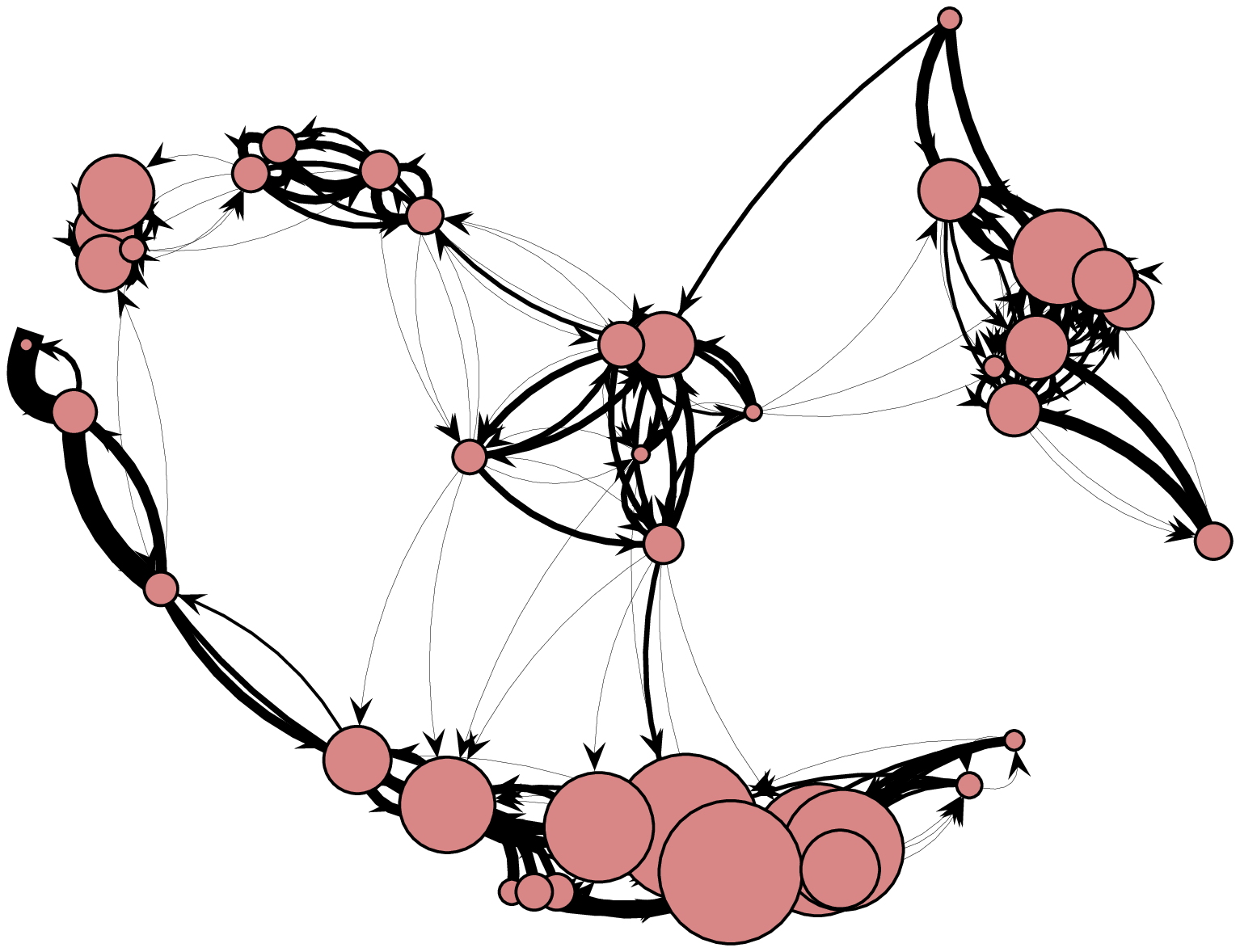}
\caption{Exemplary networks in \texttt{ariadne} for the 39 MBA Aegean sites of Fig.\ \ref{f39sites}.  The distance scale used is $D=100$km with distances based on
estimated travel times in which land travel has a friction coefficient of 3 compared to sea travel.   Vertices are proportional to the site size (0.33, 0.67 or 1.00 for small, medium and large sites).  Top right  has $J=-1.00$, $\mu=0.500$, $\kappa=1.00$ and $\lambda=4.00$.  Top left has $\lambda=2.50$ but other values unchanged.  This shows the effect of reducing the benefits of interaction on the network.  The lower right network has $J=-0.975$, $\mu=0.500$, $\kappa=0.90$ and $\lambda=4.00$.  This choice ensures top right and bottom left networks  have Hamiltonians with the linear $W_i$ coefficient (this is effectively $-4\kappa+J+\mu$ as here $F_{ii} \approx 0$) but the coefficient of the quadratic term in $W_i$ is lower in the lower network encouraging larger variations in site sizes. The lower left network has exactly the same parameter values of the top right but site 10, Akrotiri on Thera, has been removed to reflect the post-eruption geography.} \label{fariadne}
\end{figure}

To produce our networks, we find a set of values for our $N^2$ parameters $F_{ij}$ which produce an approximate minimum for $H$ of (\ref{Hariadne}) using a Monte Carlo method. Unlike the RWGM, \texttt{ariadne} is not strictly deterministic, and only discriminates between comparably optimal solutions statistically. This introduces a stochastic element which we feel is appropriate given the uncertainties in modelling such a complex system.
Of course we do not know the values for our input parameters, so our approach is to look at how the networks change as we change our parameters. Some examples are given in Fig. \ref{fariadne}. For instance reducing the benefits of interaction, reducing $\lambda$, destroys the longer distance weak links that are maintaining the global connectivity.  This also produces a general reduction in site sizes despite the maintenance of more localised networks.  Likewise, the variation in city sizes, a feature which emerges in this period and exemplified by the size of Knossos, can be produced by emphasising the non-linear terms in our Hamiltonian (also an attractive feature of the RWGM). Finally, though we produce equilibrium networks, we can tackle questions about time evolution.  Slow evolution can be simulated by comparing different values for our parameters, mimicking adiabatic changes in physical thermodynamic problems.  However we can also look at `quenches', as in removing Akrotiri after the eruption of Thera \cite{KER11}.

\section{Conclusions}

Despite the well-established nature of many of the network modelling approaches discussed here, a detailed statistical comparison remains to be given. The main purpose of this article has been to start this process by considering the relationship among popular archaeological models which take geography as their primary driving force.  Some of their key properties are summarised in table \ref{tsummary}.
\begin{table}[htb]
\begin{tabular}{l@{ }l|c@{ }c@{  }c|c|c}
       \multicolumn{2}{c|}{Model}         & \multicolumn{3}{|c}{Network Type} & \multicolumn{2}{|c}{Site Sizes}         \\
               Type      & (Fig.)          &  W &  D & Other                   & Outflows         & Inflows              \\ \hline\hline
               Voronoi   &                 & \multicolumn{3}{|c|}{none}        & \multicolumn{2}{|c}{Fixed Equal}        \\ \hline
               XTent     &                 & UW &  D & Trees                   & \multicolumn{2}{|c}{Fixed \& Different} \\ \hline
               MDN       &(\ref{fMDN})     & UW & UD &                         & \multicolumn{2}{|c}{Fixed Equal}        \\ \hline
               PPA       &(\ref{fPPA})     & UW &  D & $k_\mathrm{out}>0$      & \multicolumn{2}{|c}{Fixed Equal}        \\ \hline
               Simple GM &(\ref{fsGM})     &  W & UD &                         & Fixed Different & Fixed Different       \\ \hline
               DCGM      &(\ref{fDCGM})    &  W &  D & $k_\mathrm{in},k_\mathrm{out}>0$          & Fixed Different & Fixed Different       \\ \hline
               RWGM      &(\ref{fRWGM})    &  W &  D & $k_\mathrm{out}>0$      & Fixed Different & Variable              \\ \hline
               \texttt{ariadne}   &(\ref{fariadne}) &  W &  D &                         & Variable        & Variable              \\ \hline
\end{tabular}
\caption{Summary of some of the features of different models of interactions for geographically embedded systems. (U)W = (un)weighted, (U)D = (un)directed.  $k>0$ ($k_\mathrm{out}>0$) indicates that a model forces all sites to have at least one (outgoing) edge, however isolated it may be.} 
\label{tsummary}
\end{table}

Our emphasis has been on the last two models, the Rihll and Wilson Gravity Model (RWGM) \cite{RW87,RW91} and our own response to these issues, the \texttt{ariadne} model \cite{EKR07,KER08}.  Both of these use optimisation as a key principle, drawing on the wealth of experience from statistical physics.  However it is interesting to highlight what we consider to be positive aspects of our \texttt{ariadne} model, and how these compare with the other models discussed here:-
\begin{itemize}
  \item \texttt{ariadne} and the gravity models give weights to interactions so that we can have both strong and weak links in the nomenclature of Granovetter \cite{Gran}.
  \item \texttt{ariadne} gives the most likely arrangement of `population' over distinct sites, whereas Gravity models optimise entropy of links.
  \item \texttt{ariadne} has no absolute constraint on individual site sizes. In contrast DCGM constrains both inflow and outflow while RWGM does not constrain inflow at sites.
  \item The more isolated a site is in \texttt{ariadne}, the less it will participate.  However in gravity models and PPA, the constraints force even the most remote sites to be integrated into the system.
  \item More generally, both the RWGM and \texttt{ariadne} have to be understood statistically. From the viewpoint of ensemble theory RWGM is a microcanonical description of network flow, whereas \texttt{ariadne} provides a grand canonical description.
\end{itemize}

It can be argued that different models are capturing different types of phenomena.  For instance, the XTent and RWGM (for $\gamma >1$) are defining zones of control, which sites dominate their neighbours, rather than defining interaction patterns. In practice, the strengths of models only become apparent in specific applications to the archaeological record. Whereas PPA has been used for EBA Cyclades with rowing technology \cite{B00} we have used \texttt{ariadne} to describe the MBA S.Aegean with its sailing technology \cite{EKR07,KER08,RKE11,KER11}. The maritime networks here are characterised by an ability to make single journeys on a scale that is the typical length for which the S.Aegean is largely connected.  For networks requiring travel on a scale of several days journey (e.g. the East Mediterranean in the LBA)
the RWGM may be appropriate. We shall consider this elsewhere.

There are many properties that we have not addressed. In particular, the strongly non-linear behaviour of RWGM and \texttt{ariadne} mean that the resulting networks have a propensity for instability.   For our model these are given in more detail elsewhere \cite{EKR07,KER08,RKE11,KER11} to which we refer the reader.

\newpage
\appendix

\section{Supplementary Material}

The following tables contain all the information needed to generate the examples in this paper.  Simple replacements of text, e.g.\ using a text editor, should be able to convert the {\LaTeX} source into other suitable formats.

\begin{table}[htbp]
\begin{center}
\small
\begin{tabular}{|c|l|c|c|c|}
\hline
Number & Name & Size & Lat & Long \\ \hline
1 & Knossos & 1 & 35.3 & 25.16 \\ \hline
2 & Malia & 1 & 35.29 & 25.49 \\ \hline
3 & Phaistos & 1 & 35.05 & 24.81 \\ \hline
4 & Kommos & 0.67 & 35.02 & 24.76 \\ \hline
5 & A.Triadha & 1 & 35.06 & 24.79 \\ \hline
6 & P-kastro & 1 & 35.2 & 26.28 \\ \hline
7 & Zakros & 0.67 & 35.1 & 26.26 \\ \hline
8 & Gournia & 1 & 35.11 & 25.79 \\ \hline
9 & Chania & 1 & 35.52 & 24.02 \\ \hline
10 & Akrotiri & 0.67 & 36.35 & 25.4 \\ \hline
11 & Phylakopi & 0.67 & 36.73 & 24.42 \\ \hline
12 & Kastri & 0.67 & 36.22 & 23.06 \\ \hline
13 & Naxos & 1 & 37.11 & 25.38 \\ \hline
14 & Kea & 0.67 & 37.67 & 24.33 \\ \hline
15 & Karpathos & 0.33 & 35.42 & 27.15 \\ \hline
16 & Rhodes & 1 & 36.42 & 28.16 \\ \hline
17 & Kos & 0.67 & 36.88 & 27.28 \\ \hline
18 & Miletus & 1 & 37.84 & 27.24 \\ \hline
19 & Iasos & 0.67 & 37.28 & 27.42 \\ \hline
20 & Samos & 0.67 & 37.66 & 26.87 \\ \hline
21 & Petras & 1 & 35.2 & 26.12 \\ \hline
22 & Rethymno & 1 & 35.35 & 24.53 \\ \hline
23 & Paroikia & 0.67 & 37.08 & 25.15 \\ \hline
24 & Amorgos & 0.33 & 36.82 & 25.15 \\ \hline
25 & Ios & 0.33 & 36.73 & 25.29 \\ \hline
26 & Aegina & 0.67 & 37.75 & 23.42 \\ \hline
27 & Mycenae & 1 & 37.73 & 22.76 \\ \hline
28 & A.Stephanos & 1 & 36.8 & 22.58 \\ \hline
29 & Lavrion & 0.67 & 37.7 & 24.05 \\ \hline
30 & Kasos & 0.33 & 35.42 & 26.91 \\ \hline
31 & Kalymnos & 0.33 & 36.98 & 27.02 \\ \hline
32 & Myndus & 0.67 & 37.05 & 27.23 \\ \hline
33 & Cesme & 0.67 & 38.32 & 26.3 \\ \hline
34 & Akbuk & 0.67 & 37.41 & 27.41 \\ \hline
35 & Menelaion & 0.33 & 37.11 & 22.37 \\ \hline
36 & Argos & 0.67 & 37.63 & 22.73 \\ \hline
37 & Lerna & 0.67 & 37.5 & 22.73 \\ \hline
38 & Asine & 0.33 & 37.56 & 22.86 \\ \hline
39 & Eleusis & 0.67 & 38.04 & 23.54 \\ \hline
\end{tabular}
\end{center}
\caption{List of the positions and sizes of the 39 MBA Aegean sites.}
\label{t39sitesfull}
\end{table}

\newpage

\begin{sidewaystable}[htbp]
\begin{center}
\tiny
\setlength{\tabcolsep}{2pt}
\begin{tabular}{|c|c|c|c|c|c|c|c|c|c|c|c|c|c|c|c|c|c|c|c|c|c|c|c|c|c|c|c|c|c|c|c|c|c|c|c|c|c|c|c|}
\hline
  & \textbf{1} & \textbf{2} & \textbf{3} & \textbf{4} & \textbf{5} & \textbf{6} & \textbf{7} & \textbf{8} & \textbf{9} & \textbf{10} & \textbf{11} & \textbf{12} & \textbf{13} & \textbf{14} & \textbf{15} & \textbf{16} & \textbf{17} & \textbf{18} & \textbf{19} & \textbf{20} & \textbf{21} & \textbf{22} & \textbf{23} & \textbf{24} & \textbf{25} & \textbf{26} & \textbf{27} & \textbf{28} & \textbf{29} & \textbf{30} & \textbf{31} & \textbf{32} & \textbf{33} & \textbf{34} & \textbf{35} & \textbf{36} & \textbf{37} & \textbf{38} & \textbf{39} \\ \hline
\textbf{1} & 0 & 47 & 180 & 180 & 180 & 135 & 143 & 100 & 130 & 128 & 185 & 228 & 215 & 300 & 205 & 315 & 288 & 335 & 330 & 316 & 110 & 80 & 215 & 200 & 168 & 325 & 393 & 300 & 300 & 175 & 265 & 283 & 365 & 320 & 398 & 366 & 341 & 330 & 347 \\ \hline
\textbf{2} & 47 & 0 & 270 & 270 & 270 & 90 & 100 & 54 & 145 & 118 & 185 & 243 & 204 & 294 & 163 & 273 & 260 & 305 & 295 & 300 & 65 & 95 & 204 & 183 & 160 & 330 & 401 & 312 & 308 & 130 & 235 & 251 & 349 & 293 & 410 & 375 & 353 & 341 & 352 \\ \hline
\textbf{3} & 180 & 270 & 0 & 0 & 0 & 183 & 173 & 153 & 213 & 293 & 350 & 244 & 380 & 465 & 262 & 408 & 408 & 472 & 461 & 475 & 215 & 135 & 380 & 365 & 333 & 490 & 558 & 308 & 465 & 238 & 400 & 415 & 547 & 457 & 406 & 404 & 383 & 373 & 431 \\ \hline
\textbf{4} & 180 & 270 & 0 & 0 & 0 & 183 & 173 & 153 & 213 & 293 & 350 & 244 & 380 & 465 & 262 & 408 & 408 & 472 & 461 & 475 & 215 & 135 & 380 & 365 & 333 & 490 & 558 & 308 & 465 & 238 & 400 & 415 & 547 & 457 & 406 & 404 & 383 & 373 & 431 \\ \hline
\textbf{5} & 180 & 270 & 0 & 0 & 0 & 183 & 173 & 153 & 213 & 293 & 350 & 244 & 380 & 465 & 262 & 408 & 408 & 472 & 461 & 475 & 215 & 135 & 380 & 365 & 333 & 490 & 558 & 308 & 465 & 238 & 400 & 415 & 547 & 457 & 406 & 404 & 383 & 373 & 431 \\ \hline
\textbf{6} & 135 & 90 & 183 & 183 & 183 & 0 & 17 & 70 & 230 & 155 & 243 & 330 & 232 & 340 & 90 & 215 & 218 & 277 & 270 & 279 & 42 & 180 & 233 & 205 & 195 & 395 & 471 & 395 & 358 & 61 & 212 & 226 & 360 & 268 & 488 & 436 & 414 & 404 & 417 \\ \hline
\textbf{7} & 143 & 100 & 173 & 173 & 173 & 17 & 0 & 83 & 246 & 169 & 256 & 339 & 244 & 359 & 96 & 228 & 233 & 288 & 283 & 291 & 48 & 195 & 244 & 217 & 208 & 400 & 484 & 409 & 364 & 68 & 225 & 238 & 375 & 281 & 507 & 447 & 425 & 415 & 427 \\ \hline
\textbf{8} & 100 & 54 & 153 & 153 & 153 & 70 & 83 & 0 & 196 & 142 & 219 & 293 & 224 & 327 & 143 & 258 & 252 & 308 & 300 & 303 & 39 & 144 & 225 & 203 & 186 & 368 & 447 & 360 & 332 & 108 & 240 & 252 & 373 & 295 & 458 & 413 & 391 & 380 & 384 \\ \hline
\textbf{9} & 130 & 145 & 213 & 213 & 213 & 230 & 246 & 196 & 0 & 156 & 147 & 116 & 214 & 242 & 310 & 390 & 356 & 370 & 385 & 366 & 211 & 60 & 212 & 225 & 177 & 256 & 304 & 189 & 243 & 271 & 328 & 342 & 375 & 385 & 287 & 277 & 254 & 244 & 282 \\ \hline
\textbf{10} & 128 & 118 & 293 & 293 & 293 & 155 & 169 & 142 & 156 & 0 & 91 & 211 & 91 & 185 & 200 & 245 & 203 & 216 & 230 & 205 & 144 & 135 & 86 & 79 & 47 & 253 & 334 & 272 & 201 & 169 & 170 & 184 & 245 & 226 & 370 & 303 & 281 & 270 & 252 \\ \hline
\textbf{11} & 185 & 185 & 350 & 350 & 350 & 243 & 256 & 219 & 147 & 91 & 0 & 147 & 89 & 107 & 291 & 334 & 279 & 267 & 289 & 242 & 228 & 158 & 84 & 120 & 70 & 148 & 235 & 198 & 112 & 265 & 247 & 261 & 238 & 275 & 295 & 198 & 176 & 165 & 165 \\ \hline
\textbf{12} & 228 & 243 & 244 & 244 & 244 & 330 & 339 & 293 & 116 & 211 & 147 & 0 & 233 & 202 & 391 & 457 & 411 & 412 & 434 & 398 & 303 & 161 & 229 & 264 & 208 & 187 & 218 & 87 & 194 & 359 & 376 & 390 & 372 & 429 & 185 & 184 & 162 & 124 & 212 \\ \hline
\textbf{13} & 215 & 204 & 380 & 380 & 380 & 232 & 244 & 224 & 214 & 91 & 89 & 233 & 0 & 117 & 234 & 293 & 188 & 185 & 207 & 152 & 226 & 209 & 13 & 67 & 53 & 198 & 305 & 289 & 140 & 247 & 170 & 184 & 162 & 195 & 387 & 270 & 248 & 237 & 202 \\ \hline
\textbf{14} & 300 & 294 & 465 & 465 & 465 & 340 & 359 & 327 & 242 & 185 & 107 & 202 & 117 & 0 & 377 & 378 & 291 & 266 & 302 & 235 & 324 & 260 & 119 & 175 & 145 & 90 & 223 & 247 & 28 & 347 & 262 & 281 & 195 & 286 & 340 & 190 & 168 & 157 & 91 \\ \hline
\textbf{15} & 205 & 163 & 262 & 262 & 262 & 90 & 96 & 143 & 310 & 200 & 291 & 391 & 234 & 377 & 0 & 147 & 175 & 260 & 250 & 265 & 108 & 253 & 263 & 229 & 232 & 433 & 523 & 455 & 397 & 30 & 193 & 196 & 353 & 251 & 548 & 487 & 465 & 454 & 460 \\ \hline
\textbf{16} & 315 & 273 & 408 & 408 & 408 & 215 & 228 & 258 & 390 & 245 & 334 & 457 & 293 & 378 & 147 & 0 & 106 & 189 & 179 & 202 & 228 & 351 & 287 & 244 & 265 & 462 & 559 & 518 & 402 & 162 & 133 & 123 & 298 & 170 & 611 & 528 & 506 & 495 & 482 \\ \hline
\textbf{17} & 288 & 260 & 408 & 408 & 408 & 218 & 233 & 252 & 356 & 203 & 279 & 411 & 188 & 291 & 175 & 106 & 0 & 78 & 66 & 94 & 227 & 329 & 197 & 140 & 191 & 367 & 477 & 446 & 314 & 175 & 35 & 17 & 191 & 63 & 539 & 470 & 448 & 437 & 365 \\ \hline
\textbf{18} & 335 & 305 & 472 & 472 & 472 & 277 & 288 & 308 & 370 & 216 & 267 & 412 & 185 & 266 & 260 & 189 & 78 & 0 & 70 & 37 & 282 & 346 & 188 & 148 & 215 & 356 & 471 & 459 & 297 & 247 & 69 & 59 & 146 & 50 & 552 & 447 & 425 & 414 & 348 \\ \hline
\textbf{19} & 330 & 295 & 461 & 461 & 461 & 270 & 283 & 300 & 385 & 230 & 289 & 434 & 207 & 302 & 250 & 179 & 66 & 70 & 0 & 88 & 276 & 352 & 215 & 176 & 230 & 386 & 501 & 490 & 332 & 240 & 65 & 49 & 197 & 41 & 583 & 453 & 431 & 420 & 384 \\ \hline
\textbf{20} & 316 & 300 & 475 & 475 & 475 & 279 & 291 & 303 & 366 & 205 & 242 & 398 & 152 & 235 & 265 & 202 & 94 & 37 & 88 & 0 & 283 & 332 & 163 & 132 & 194 & 335 & 443 & 442 & 275 & 252 & 81 & 76 & 117 & 70 & 535 & 407 & 385 & 374 & 336 \\ \hline
\textbf{21} & 110 & 65 & 215 & 215 & 215 & 42 & 48 & 39 & 211 & 144 & 228 & 303 & 226 & 324 & 108 & 228 & 227 & 282 & 276 & 283 & 0 & 157 & 223 & 197 & 185 & 375 & 452 & 372 & 337 & 76 & 216 & 230 & 353 & 272 & 465 & 424 & 402 & 391 & 388 \\ \hline
\textbf{22} & 80 & 95 & 135 & 135 & 135 & 180 & 195 & 144 & 60 & 135 & 158 & 161 & 209 & 260 & 253 & 351 & 329 & 346 & 352 & 332 & 157 & 0 & 201 & 206 & 164 & 281 & 341 & 235 & 263 & 220 & 296 & 308 & 368 & 349 & 328 & 313 & 290 & 280 & 309 \\ \hline
\textbf{23} & 215 & 204 & 380 & 380 & 380 & 233 & 244 & 225 & 212 & 86 & 84 & 229 & 13 & 119 & 263 & 287 & 197 & 188 & 215 & 163 & 223 & 201 & 0 & 62 & 44 & 210 & 310 & 282 & 146 & 239 & 188 & 203 & 168 & 206 & 375 & 253 & 231 & 220 & 185 \\ \hline
\textbf{24} & 200 & 183 & 365 & 365 & 365 & 205 & 217 & 203 & 225 & 79 & 120 & 264 & 67 & 175 & 229 & 244 & 140 & 148 & 176 & 132 & 197 & 206 & 62 & 0 & 67 & 254 & 349 & 315 & 203 & 199 & 124 & 141 & 175 & 164 & 408 & 257 & 235 & 224 & 197 \\ \hline
\textbf{25} & 168 & 160 & 333 & 333 & 333 & 195 & 208 & 186 & 177 & 47 & 70 & 208 & 53 & 145 & 232 & 265 & 191 & 215 & 230 & 194 & 185 & 164 & 44 & 67 & 0 & 207 & 302 & 266 & 159 & 204 & 177 & 192 & 211 & 229 & 359 & 259 & 249 & 250 & 212 \\ \hline
\textbf{26} & 325 & 330 & 490 & 490 & 490 & 395 & 400 & 368 & 256 & 253 & 148 & 187 & 198 & 90 & 433 & 462 & 367 & 356 & 386 & 335 & 375 & 281 & 210 & 254 & 207 & 0 & 170 & 237 & 77 & 410 & 356 & 371 & 289 & 383 & 330 & 164 & 142 & 131 & 38 \\ \hline
\textbf{27} & 393 & 401 & 558 & 558 & 558 & 471 & 484 & 447 & 304 & 334 & 235 & 218 & 305 & 223 & 523 & 559 & 477 & 471 & 501 & 443 & 452 & 341 & 310 & 349 & 302 & 170 & 0 & 268 & 209 & 493 & 467 & 481 & 410 & 492 & 330 & 30 & 60 & 75 & 190 \\ \hline
\textbf{28} & 300 & 312 & 308 & 308 & 308 & 395 & 409 & 360 & 189 & 272 & 198 & 87 & 289 & 247 & 455 & 518 & 446 & 459 & 490 & 442 & 372 & 235 & 282 & 315 & 266 & 237 & 268 & 0 & 241 & 424 & 429 & 445 & 423 & 472 & 93 & 233 & 209 & 212 & 246 \\ \hline
\textbf{29} & 300 & 308 & 465 & 465 & 465 & 358 & 364 & 332 & 243 & 201 & 112 & 194 & 140 & 28 & 397 & 402 & 314 & 297 & 332 & 275 & 337 & 263 & 146 & 203 & 159 & 77 & 209 & 241 & 0 & 369 & 290 & 309 & 214 & 314 & 334 & 174 & 152 & 141 & 77 \\ \hline
\textbf{30} & 175 & 130 & 238 & 238 & 238 & 61 & 68 & 108 & 271 & 169 & 265 & 359 & 247 & 347 & 30 & 162 & 175 & 247 & 240 & 252 & 76 & 220 & 239 & 199 & 204 & 410 & 493 & 424 & 369 & 0 & 179 & 194 & 332 & 236 & 517 & 468 & 446 & 435 & 418 \\ \hline
\textbf{31} & 265 & 235 & 400 & 400 & 400 & 212 & 225 & 240 & 328 & 170 & 247 & 376 & 170 & 262 & 193 & 133 & 35 & 69 & 65 & 81 & 216 & 296 & 188 & 124 & 177 & 356 & 467 & 429 & 290 & 179 & 0 & 19 & 175 & 58 & 522 & 426 & 404 & 393 & 353 \\ \hline
\textbf{32} & 283 & 251 & 415 & 415 & 415 & 226 & 238 & 252 & 342 & 184 & 261 & 390 & 184 & 281 & 196 & 123 & 17 & 59 & 49 & 76 & 230 & 308 & 203 & 141 & 192 & 371 & 481 & 445 & 309 & 194 & 19 & 0 & 175 & 47 & 538 & 438 & 416 & 405 & 366 \\ \hline
\textbf{33} & 365 & 349 & 547 & 547 & 547 & 360 & 375 & 373 & 375 & 245 & 238 & 372 & 162 & 195 & 353 & 298 & 191 & 146 & 197 & 117 & 353 & 368 & 168 & 175 & 211 & 289 & 410 & 423 & 214 & 332 & 175 & 175 & 0 & 184 & 516 & 376 & 354 & 343 & 295 \\ \hline
\textbf{34} & 320 & 293 & 457 & 457 & 457 & 268 & 281 & 295 & 385 & 226 & 275 & 429 & 195 & 286 & 251 & 170 & 63 & 50 & 41 & 70 & 272 & 349 & 206 & 164 & 229 & 383 & 492 & 472 & 314 & 236 & 58 & 47 & 184 & 0 & 565 & 459 & 437 & 425 & 387 \\ \hline
\textbf{35} & 398 & 410 & 406 & 406 & 406 & 488 & 507 & 458 & 287 & 370 & 295 & 185 & 387 & 340 & 548 & 611 & 539 & 552 & 583 & 535 & 465 & 328 & 375 & 408 & 359 & 330 & 330 & 93 & 334 & 517 & 522 & 538 & 516 & 565 & 0 & 300 & 270 & 327 & 339 \\ \hline
\textbf{36} & 366 & 375 & 404 & 404 & 404 & 436 & 447 & 413 & 277 & 303 & 198 & 184 & 270 & 190 & 487 & 528 & 470 & 447 & 453 & 407 & 424 & 313 & 253 & 257 & 259 & 164 & 30 & 233 & 174 & 468 & 426 & 438 & 376 & 459 & 300 & 0 & 30 & 54 & 220 \\ \hline
\textbf{37} & 341 & 353 & 383 & 383 & 383 & 414 & 425 & 391 & 254 & 281 & 176 & 162 & 248 & 168 & 465 & 506 & 448 & 425 & 431 & 385 & 402 & 290 & 231 & 235 & 249 & 142 & 60 & 209 & 152 & 446 & 404 & 416 & 354 & 437 & 270 & 30 & 0 & 15 & 158 \\ \hline
\textbf{38} & 330 & 341 & 373 & 373 & 373 & 404 & 415 & 380 & 244 & 270 & 165 & 124 & 237 & 157 & 454 & 495 & 437 & 414 & 420 & 374 & 391 & 280 & 220 & 224 & 250 & 131 & 75 & 212 & 141 & 435 & 393 & 405 & 343 & 425 & 327 & 54 & 15 & 0 & 152 \\ \hline
\textbf{39} & 347 & 352 & 431 & 431 & 431 & 417 & 427 & 384 & 282 & 252 & 165 & 212 & 202 & 91 & 460 & 482 & 365 & 348 & 384 & 336 & 388 & 309 & 185 & 197 & 212 & 38 & 190 & 246 & 77 & 418 & 353 & 366 & 295 & 387 & 339 & 220 & 158 & 152 & 0 \\ \hline
\end{tabular}
\end{center}
\caption{Distances between the 39 MBA Aegean sites of Table \ref{f39sites} used in this paper.  The column and row labels refer to the site numbers in Table \ref{t39sites}. The distances are in units of km using estimated best routes for minimum travel time.  Land travel has a friction coefficient of 3 compared to sea travel.  }
\label{tdistances}
\end{sidewaystable}


\begin{thebibliography}{10}
\providecommand{\urlprefix}{}
\expandafter\ifx\csname urlstyle\endcsname\relax
  \providecommand{\doi}[1]{doi:\discretionary{}{}{}#1}\else
  \providecommand{\doi}{doi:\discretionary{}{}{}\begingroup
  \urlstyle{rm}\Url}\fi

\bibitem{BBSW05}
Balister, P., Bollob\'{a}s, B., Sarkar, A., and Walters, M., Connectivity of
  random k-nearest-neighbour graphs, \emph{Adv. in Appl. Probab.} \textbf{37}
  (2005) 1--24.

\bibitem{B08a}
Bevan, A., Computational models for understanding movement and territory, in
  \emph{Sistemas de Informaci\'{o}n Geogr\'{a}fica y An\'{a}lisis Arquel\'{o}gico del
  Territorio. V Simposio Internacional de Arqueolog\'{i}a de M\'{e}rida, Anejos de
  Archivo Espa\~{n}ol de Arqueolog\'{i}a.}, eds. Mayoral, V. and Celestino, S. (2008).

\bibitem{B10c}
Bevan, A., Political geography and palatial Crete, \emph{J.Mediterranean Arch.} \textbf{23} (2010).

\bibitem{B00}
Broodbank, C., \emph{An Island Archaeology of the Early Cyclades} (CUP, 2000).

\bibitem{B10}
Barthel\'{e}my, M., Spatial Networks, Physics Reports \textbf{499} (2011) 1--101  .


\bibitem{CW85}
Clarke, M. and Wilson, A.~G., The dynamics of urban spatial structure: the
  progress of a research programme, \emph{Trans.\ Inst.\ Br.\ Geogr.\ New Ser.}
  \textbf{10} (1985) 427–451.

\bibitem{C07}
Collar, A., Network theory and religious innovation, \emph{Mediterranean
  Historical Review} \textbf{22} (2007) 149--162.

\bibitem{DW09}
Dearden, J. and Wilson, A., Exploring urban retail phase transitions, Technical
  report, UCL (2009).

\bibitem{DW10}
  Dearden, J., and Wilson, A.,
  Urban retail phase transitions – 1: an analysis system,
  Technical report, UCl (2010).

\bibitem{ES90}
  Erlander, S., and Stewart, N.F.,
  \emph{The Gravity Model in Transportation Analysis}
  (VSP, 1990).

\bibitem{E10}
Evans, T.S., Clique graphs and overlapping communities, \emph{J.Stat.Mech}
  (2010) P12037.

\bibitem{EKR07}
Evans, T., Knappett, C., and Rivers, R., Using statistical physics to
  understand relational space: A case study from mediterranean prehistory, in
  \emph{Complexity Perspectives on Innovation and Social Change}, eds. Lane,
  D., Pumain, D., van~der Leeuw, S., and West, G., \emph{Methodos Series},
  Vol.~7, chapter~17 (Springer, 2009), pp. 451--479.

\bibitem{EL09}
 Evans, T.S., and Lambiotte, R..
 Line Graphs, Link Partitions and Overlapping Communities,
 \emph{Phys.Rev.E} \textbf{80} (2009) 016105.

\bibitem{Gran}
 Granovetter, M. S., The Strength of Weak Ties,
 \emph{Am.J.Sociology} \textbf{78} (1973) 1360.

\bibitem{HH91}
Hage, P. and Harary, F. \emph{Exchange in Oceania: a graph theoretic analysis.}  Oxford, Clarendon Press 1991.

\bibitem{I83}
Irwin, G. 1983. Chieftainship, kula and trade in Massim prehistory, in Leach, J.W. and Leach, E. (eds.)
\emph{The Kula: new perspectives on Massim exchange:} 29--72. Cambridge: Cambridge University Press.

\bibitem{KER08}
Knappett, C., Evans, T., and Rivers, R., Modelling maritime interaction in the
  aegean bronze age, \emph{Antiquity} \textbf{82} (2008) 1009--1024.

\bibitem{KER11}
Knappett, C., Evans, T., and Rivers, R., The Theran eruption and Minoan palatial collapse: new interpretations gained from
modelling the maritime network, \emph{Antiquity} (in press) (2011).

\bibitem{KGR05}
Kohler, T.~A., Gumerman, G.~J., and Reynolds, R.~G., Simulating ancient
  societies, \emph{Scientific American}  (2005).

\bibitem{OW94}
   Ort\'{u}zar, J. de D., and Willumsen, L.G.
  \emph{Modelling Transport},
  (Wiley, 1994)


\bibitem{PS04}
 Poduri, S. and Sukhatme, G.~S.,
 Constrained coverage for mobile sensor networks,
 in \emph{IEEE International Conference on Robotics and Automation} (2004), pp. 165--172.


\bibitem{P03}
 Penrose, M.,
 \emph{Random Geometric Graphs} (OUP, 2003).


\bibitem{R75}
Renfrew, C., Trade as action at a distance, in \emph{Ancient Civilization and
  Trade}, eds. Sabloff, J. and Lamberg-Karlovsky, C. (University of New Mexico
  Press, Albuquerque), 1975), pp. 3--59.

\bibitem{RL79}
Renfrew, C. and Level, E., Exploring dominance: predicting polities from
  centres, in \emph{Transformations: Mathematical Approaches to Culture
  Change}, eds. Renfrew, A. and Cooke, K. (Academic Press, London, 1979), pp.
  145--67.

\bibitem{RW87}
Rihll, T.~E. and Wilson, A.~G., Spatial interaction and structural models in
  historical analysis: Some possibilities and an example, \emph{Histoire \&
  Mesure} \textbf{2} (1987) 5--32.

\bibitem{RW91}
Rihll, T. and Wilson, A., Modelling settlement structures in ancient greece:
  new approaches to the polis, in \emph{City and country in the ancient world},
  eds. Rich, J. and Wallace-Hadrill, A., chapter~3 (1991).

\bibitem{RKE11}
Rivers, R., Knappett, C., and Evans, T., Network models and archaeological
  spaces, in \emph{Computational Approaches to Archaeological Spaces}, eds.
  A.Bevan and Lake, M. (Left Coast Press, 2011).

\bibitem{Sind07}
Sindb{\ae}k, S.~M., Networks and nodal points: the emergence of towns in early
  Viking age Scandinavia, \emph{Antiquity} \textbf{81} (2007) 119–132.

\bibitem{S07b}
Sindb{\ae}k, S.~M., The small world of the Vikings: Networks in early medieval
  communication and exchange, \emph{Norwegian Archaeological Review}
  \textbf{40} (2007) 59--74.

\bibitem{SH10}
Srinivasa, S. and Haenggi, M., Distance distributions in finite uniformly
  random networks: Theory and applications, \emph{IEEE Transactions on
  Vehicular Technology} \textbf{59} (2010) 940--949.

\bibitem{T77}
Terrell, J., Human biogeography in the solomon islands, \emph{Fieldiana,
  Anthropology} \textbf{68} (1977).

\bibitem{T10}
Terrell, J.~E., Language and material culture on the Sepik coast of Papua New
  Guinea: Using social network analysis to simulate, graph, identify, and
  analyze social and cultural boundaries between communities, \emph{The Journal
  of Island and Coastal Archaeology} \textbf{5} (2010) 3--32.


\bibitem{WGCWSKWSS07}
Wilkinson, T.~J., Gibson, M., Christiansen, J.~H., Widell, M., Schloen, D.,
  Kouchoukos, N., Woods, C., Sanders, J., Simunich, K.-L., Altaweel, M., Ur,
  J.~A., Hritz, C., Lauinger, J., Paulette, T., and Tenney, J., Modeling
  settlement systems in a dynamic environment: Case studies from mesopotamia,
  in \emph{Model Based Archaeology}, eds. Kohler, T. and van~der Leeuw, S.~E.
  (Santa Fe, New Mexico: School for Advanced Research Press, 2007), pp.\
  175--208, ISSN/ISBN: 1-930618-87-9.

\bibitem{W08}
Wilson, A., Boltzmann, Lotka and Volterra and spatial structural evolution: an
  integrated methodology for some dynamical systems, \emph{J.R.Soc.Interface} \textbf{5} (2008) 865--871.

\bibitem{WD10}
Wilson, A. and Dearden, J., Phase transitions and path dependence in urban
  evolution, \emph{Journal of Geographical Systems}  (2010) 1--16.


\end{thebibliography}
\end{document}